\documentclass{article} 
\usepackage{iclr2027_conference,times}

\usepackage{amsmath,amsfonts,bm}

\def\eqref#1{equation~\ref{#1}}

\def\1{\bm{1}}

\DeclareMathAlphabet{\mathsfit}{\encodingdefault}{\sfdefault}{m}{sl}
\SetMathAlphabet{\mathsfit}{bold}{\encodingdefault}{\sfdefault}{bx}{n}

\usepackage{hyperref}
\usepackage{url}
\usepackage{graphicx}
\usepackage{booktabs}
\hypersetup{hidelinks}

\title{Audio Tokens as a Budgeted Resource: Marginal-Utility Allocation for Scalable Audio Representations}

\author{
Mingyu Zhao$^{1}$ \quad
Jinchao Zhang$^{2,*}$ \quad
Zhiyong Wu$^{1,*}$\\
$^{1}$Tsinghua Shenzhen International Graduate School, Tsinghua University\\
$^{2}$Tencent\\
\texttt{zmy24@mails.tsinghua.edu.cn}\\
\texttt{dayerzhang@tencent.com} \quad
\texttt{zywu@sz.tsinghua.edu.cn}\\
$^{*}$ Corresponding authors.
}

\iclrfinalcopy 
\begin{document}

\maketitle

\begin{abstract}
Discrete audio tokens are widely used as a representation interface, yet fixed-depth RVQ tokenizers allocate equal capacity to every frame despite varying refinement value. We introduce UniAdapt, which learns marginal utility of RVQ refinements on a frozen codec and allocates them under exact serialized-bit budgets. A rate-independent causal controller predicts acoustic utility, while an optional semantic head supports speech-only utterance-level allocation; measured acoustic and semantic marginal gains on speech have a correlation of 0.42. For causal allocation, a primal–dual allocator selects prefix-valid depths, while an exact guard constrains each sequence prefix to its matched fixed-depth serialized budget. Under utterance-level allocation, UniAdapt reduces Log-STFT distortion by 1.07–4.39\% across speech, music, and environmental audio without larger budgets. Causally, it improves three of four speech rates with zero violations across 800 utterance--rate evaluations and runs faster than real time. A 20-listener utterance-level MUSHRA study shows a significant 3.52-point speech improvement, with no significant differences on music or environmental audio. These results support separating utility prediction from budget enforcement for scalable, budget-conditioned audio representations.
\end{abstract}

\section{Introduction}
\label{sec:introduction}

Discrete audio tokens are increasingly used as a common interface for speech
and audio generation, understanding, and interactive systems
\citep{zhang2024speechtokenizer,huang2024repcodec,ji2025wavtokenizer,
defossez2024moshi}. Residual vector quantization (RVQ) is particularly
attractive because it exposes a hierarchy of discrete refinements
\citep{zeghidour2022soundstream,defossez2023encodec,kumar2023dac}.
Yet fixed-depth RVQ activates the same number of layers at every frame for a
given operating point, implicitly allocating equal capacity over time. Real
audio violates this assumption: an extra refinement can be valuable near a
phonetic transition or transient event but largely redundant in a stationary
region. This raises a simple question: \emph{given a limited representation
budget, where is the next RVQ refinement worth spending?}

Existing work addresses different parts of this problem. Variable-rate and
content-adaptive codecs relax uniform allocation
\citep{chae2025vrvq,zhang2025variableframerate,li2026flexicodec}, while
semantic-aware tokenizers show that acoustic fidelity and semantic
preservation need not favor the same representation
\citep{zhang2024speechtokenizer,huang2024repcodec,li2025dualcodec,
zhao2026spgcodec}. Most closely related, BAMU
\citep{zhao2026bamu} already introduced rate-independent marginal-utility
prediction and exact realized-container budget enforcement for frozen speech
codecs, using a non-causal predictor and full-utterance allocation under an exact serialized-size budget. SPG-Codec \citep{zhao2026spgcodec} showed that the benefit of frozen semantic
priors depends strongly on bitrate, but incorporated semantics during codec
training rather than post-hoc token allocation. UniStream
\citep{zhao2026unistream} provides the frozen causal 48-kHz RVQ hierarchy used
here for speech, music, and environmental audio, but does not determine how its
capacity should vary over time. Building on these directions, UniAdapt focuses
on causal depth decisions under exact per-prefix budgets and on
objective-dependent allocation over the same frozen hierarchy. This distinction
matters in streaming: decisions are irreversible, while the true transmitted
cost includes code indices, depth maps, expert side information, headers,
run-length coding, and byte padding, so an utterance-level budget does not
ensure prefix feasibility.

Our central idea is to treat audio tokens as a \emph{budgeted resource}:
predict token value independently of the target rate, choose the representation
objective, and introduce the resource constraint only when capacity is
allocated. Rather than predicting a bitrate-specific depth, UniAdapt estimates
the marginal utility of activating one additional RVQ refinement at each frame
and layer. On speech, measured acoustic and HuBERT-based semantic marginal
gains have an overall correlation of $0.4205$, falling to near zero or slightly
negative values in deeper layers, which motivates separate objective-dependent
utility models. The causal streaming system uses acoustic utility only; an
optional semantic head is used for speech-only, utterance-level
objective-dependent allocation.

We instantiate this principle on a frozen causal universal audio codec. A
lightweight causal controller predicts frame- and layer-wise acoustic marginal
utilities without receiving the target rate. For online operation, a causal
primal--dual allocator converts these utilities into prefix-valid RVQ depths
under token-bucket and switching controls. Before committing each depth, an
exact feasibility guard evaluates the realized serialized prefix cost against
the matched fixed-depth prefix budget and additionally checks that the next
frame can still transmit the mandatory minimum depth $d=1$. Depth and expert
maps, the header, and byte padding are all included in the accounting. The
committed depths define active prefixes of the same frozen RVQ hierarchy,
yielding a fully decodable dynamic representation with the same codebooks and
decoder rather than separately trained rate-specific tokenizers.

Experiments keep utterance-level and causal settings separate. Under
utterance-level exact-budget allocation, one three-domain acoustic controller
reduces Log-STFT distortion by $1.07$--$4.39\%$ across speech, music, and
environmental audio without exceeding matched fixed-depth budgets. In the
causal setting, UniAdapt improves three of four speech rates, produces zero
budget violations across $800$ utterance--rate evaluations, and runs complete
waveform-to-waveform streaming faster than real time. A $20$-listener MUSHRA
study using utterance-level allocation shows a significant $3.52$-point speech
improvement, with no significant difference on music or environmental audio.
Frozen-token evaluations show significant CER reductions under a shared
multi-rate CTC probe, higher FSD50K macro-mAP, and lower speaker-verification
EER at depths 3, 4, and 6. At lower realized bitrates, the main MTG-Jamendo
tagging and Clotho mean-recall metrics show no statistically significant
differences. On LibriSpeech, dynamic allocation improves PESQ and STOI over the
same frozen backbone at a slightly lower rate, although EnCodec retains higher
absolute reconstruction quality at a similar bitrate. These results support
content-adaptive allocation under explicit serialized-bit constraints rather
than a claim of reconstruction state of the art.

\paragraph{Contributions.}
Our contributions are fourfold:
\begin{itemize}
    \item We extend marginal-utility allocation on frozen RVQ hierarchies from
    utterance-level optimization to \textbf{causal, budget-conditioned token
    allocation}, using a rate-independent causal predictor across operating
    rates.

    \item We model \textbf{objective-dependent marginal utility} on the same
    frozen hierarchy, with acoustic utility for causal operation and optional
    semantic utility for speech-only utterance-level allocation.

    \item We develop a \textbf{causal exact-budget allocator} combining
    primal--dual rate control with serialized per-prefix feasibility checks,
    keeping each committed prefix within its matched fixed-depth budget while
    ensuring next-frame minimum-depth feasibility.

    \item We validate the resulting \textbf{budget-conditioned frozen-token
    representation} across multiple rates and audio domains, downstream tasks,
    human listening, and real-time waveform streaming.
\end{itemize}

\section{Related Work}
\label{sec:related_work}

\subsection{Neural Audio Codecs and Adaptive Tokenization}

RVQ-based codecs such as SoundStream~\citep{zeghidour2022soundstream},
EnCodec~\citep{defossez2023encodec}, and DAC~\citep{kumar2023dac}
establish hierarchical residual quantization for neural audio coding.
Related codec architectures explore multi-scale quantization, low token rates,
and causal universal audio coding
\citep{siuzdak2024snac,defossez2024moshi,ji2025wavtokenizer,
zhao2026unistream}. These methods primarily develop representation
architectures and their supported operating points. UniAdapt instead studies
how existing RVQ refinements should be allocated across frames under explicit
serialized-bit budgets. Adaptive methods vary token usage through active RVQ
depth or temporal resolution
\citep{chae2025vrvq,zhang2025variableframerate,li2026flexicodec}.
BAMU~\citep{zhao2026bamu} is most closely related: it learns
rate-independent acoustic marginal utilities on frozen speech codecs using a
non-causal predictor and full-utterance allocation under an exact
serialized-size budget. UniAdapt
extends this formulation to causal, irreversible per-frame decisions under
per-prefix budgets, while separately studying objective-dependent utility.
UniStream~\citep{zhao2026unistream} supplies the causal universal RVQ
hierarchy used here but does not address content-adaptive per-frame depth.

\subsection{Semantic-Aware Audio Representations}

Semantic-aware tokenizers incorporate linguistic information through different
mechanisms. SpeechTokenizer~\citep{zhang2024speechtokenizer} organizes
semantic and acoustic information across RVQ layers.
RepCodec~\citep{huang2024repcodec} learns discrete semantic tokens by
reconstructing continuous representations from pretrained speech encoders,
whereas DualCodec~\citep{li2025dualcodec} integrates self-supervised speech
features and waveform representations through dual-stream encoding.
SPG-Codec~\citep{zhao2026spgcodec} further shows that the usefulness of frozen
semantic priors varies with bitrate, but incorporates semantic constraints
during codec training. UniAdapt instead keeps the codec and RVQ hierarchy
frozen and uses semantic preservation only to define the marginal value of
transmitting an additional existing refinement. This semantic branch is
restricted to speech and utterance-level allocation; causal streaming uses
acoustic utility only. UniAdapt therefore studies objective-dependent
allocation rather than semantic disentanglement.

\subsection{Causal and Exact-Budget Allocation}

Streaming turns adaptive tokenization into an irreversible online
resource-allocation problem. Nominal bitrate or average RVQ depth alone does
not determine the exact serialized size. In the dynamic container used here,
the cost includes active code indices, run-length-coded depth and expert maps,
the header, and byte padding. BAMU~\citep{zhao2026bamu} enforces an exact
utterance-level container budget, but its full-sequence optimization does not
constrain every intermediate prefix. UniAdapt instead requires each committed
prefix to remain within its matched fixed-depth serialized budget while
retaining minimum-depth $d=1$ feasibility for the next frame.

Together, these lines motivate allocation over a frozen RVQ hierarchy with
causal acoustic-utility prediction and exact per-prefix budget enforcement,
alongside an offline speech setting for objective-dependent acoustic--semantic
allocation. UniAdapt targets this intersection rather than a new
reconstruction backbone.

\section{Method}
\label{sec:method}

\subsection{Frozen Hierarchy and Problem Formulation}
\label{sec:formulation}

\begin{figure}[t]
    \centering
    \includegraphics[width=\linewidth]{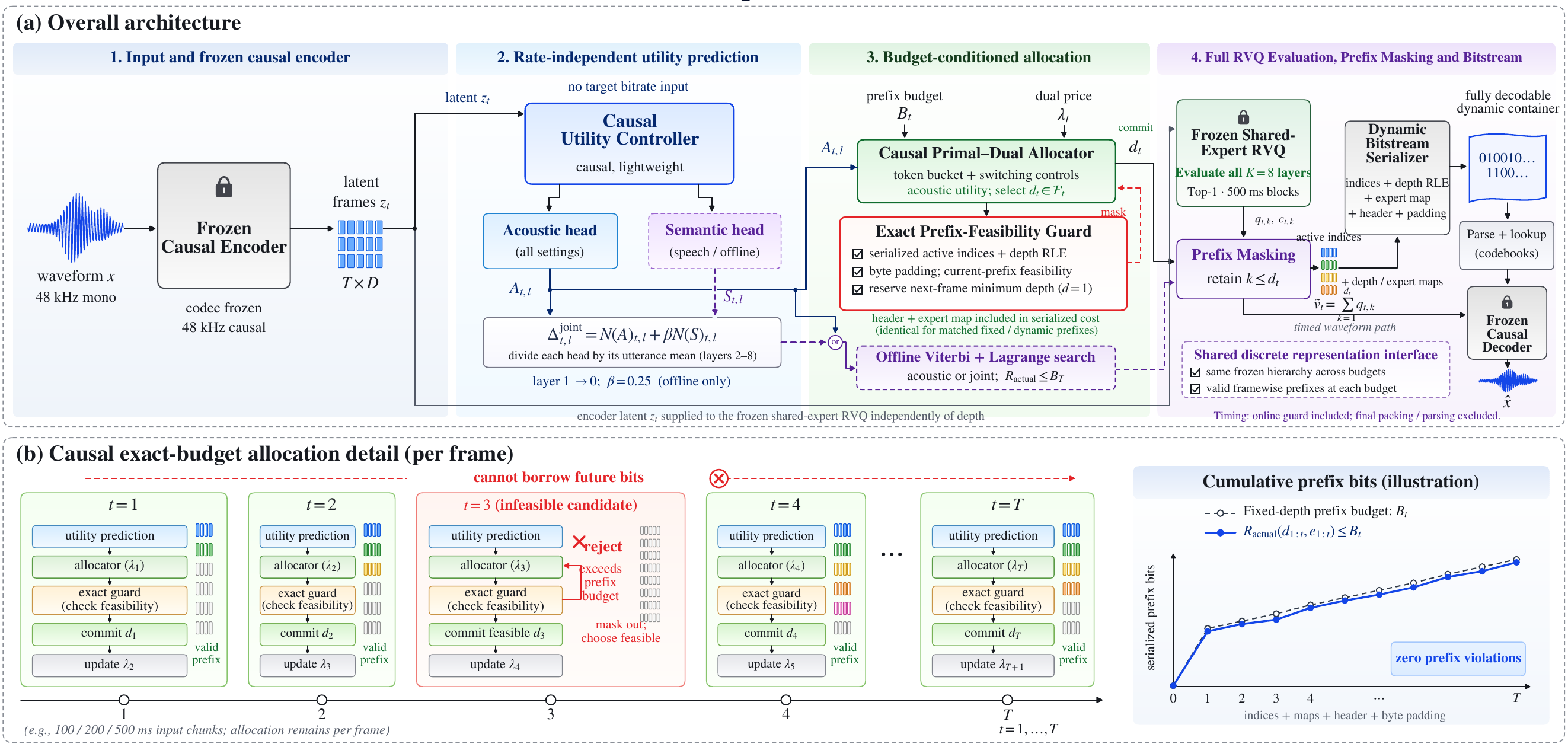}
    \caption{
    \textbf{UniAdapt framework.}
    (a) A frozen causal RVQ codec provides a shared hierarchy of audio
    refinements. A rate-independent causal controller predicts acoustic
    marginal utilities for all settings, with an optional semantic head for
    speech-only utterance-level allocation. A budget-conditioned allocator
    selects framewise prefix depths under exact serialized-prefix feasibility.
    The reported real-backbone path evaluates all RVQ layers and applies the
    selected prefix mask before serialization and frozen causal decoding.
    (b) Causal allocation is irreversible: each committed prefix must satisfy
    its matched fixed-depth serialized budget, while reserving minimum-depth
    feasibility for the next frame.
    }
    \label{fig:framework}
\end{figure}

As shown in Fig.~\ref{fig:framework}(a), UniAdapt operates on a frozen causal
RVQ codec. The encoder produces latent frames
\begin{equation}
z=E(x)=\{z_t\}_{t=1}^{T},\qquad z_t\in\mathbb{R}^{C},
\tag{1}
\end{equation}
and the frozen quantizer input projection initializes the residual,
\begin{equation}
v_t=P_{\mathrm{in}}(z_t)\in\mathbb{R}^{C_q},\qquad
r_{t,0}=v_t,\qquad r_{t,k}=r_{t,k-1}-q_{t,k},
\tag{2}
\end{equation}
where $C_q$ is the quantization-space dimensionality. UniAdapt selects
$d_t\in\{1,\ldots,K\}$ and uses
$\tilde v_t(d_t)=\sum_{k=1}^{d_t}q_{t,k}$. The encoder, input projection,
codebooks, routing modules, and decoder remain frozen. We use the reversible
Top-1 shared-expert mode: expert assignments are derived from the frozen
router, shared across acoustic RVQ layers within fixed temporal blocks, and
serialized rather than optimized by the depth allocator.

Let $R_{\mathrm{actual}}(d_{1:t},e_{1:t})$ denote the size obtained by
serializing a sequence prefix, including active indices, depth and expert
maps, the header, and byte padding. For matched fixed depth $D$, using the
same expert map,
\begin{equation}
B_t=R_{\mathrm{actual}}^{\mathrm{fixed}}(D,t).
\tag{3}
\end{equation}
We consider
\begin{equation}
\max_{\{d_t\}}\sum_{t=1}^{T}U_t(d_t)
-\rho\sum_{t=2}^{T}\mathbf{1}[d_t\neq d_{t-1}]
\quad\mathrm{s.t.}\quad
R_{\mathrm{actual}}(d_{1:\tau},e_{1:\tau})\le B_\tau,\ \forall\tau .
\tag{4}
\end{equation}
Equation~(4) defines the constrained objective. Given the supplied budget
reference and target rate, online decisions depend only on $z_{\le t}$ and
persistent allocator state; the utility predictor itself receives no
target-rate input.

\subsection{Rate-Independent Acoustic and Semantic Marginal Utility}
\label{sec:utility}

The controller predicts refinement value rather than bitrate-specific depth.
Acoustic utility is measured in the projected RVQ space:
\begin{equation}
D^{\mathrm{ac}}_{t,k}=\frac{1}{C_q}\|r_{t,k}\|_2^2,\qquad
\Delta^{\mathrm{ac}}_{t,k}
=\left[D^{\mathrm{ac}}_{t,k-1}-D^{\mathrm{ac}}_{t,k}\right]_+,
\quad k=1,\ldots,K.
\tag{5}
\end{equation}

For speech, a frozen HuBERT teacher $\Phi$ provides semantic targets. Let
$\hat x^{(k)}$ be reconstruction at uniform depth $k$, and let $j$ index the
HuBERT frames:
\begin{equation}
D^{\mathrm{sem}}_{j,k}
=1-\operatorname{cos}\!\left(\Phi_j(x),\Phi_j(\hat x^{(k)})\right).
\tag{6}
\end{equation}
For $k\ge2$, differencing and clipping precede linear interpolation to the
codec grid:
\begin{equation}
\Delta^{\mathrm{sem}}_{t,k}
=\mathcal I_{J\rightarrow T}
\left(\left[D^{\mathrm{sem}}_{\cdot,k-1}
-D^{\mathrm{sem}}_{\cdot,k}\right]_+\right)_t,
\qquad \Delta^{\mathrm{sem}}_{t,1}=0.
\tag{7}
\end{equation}
Here $\mathcal I$ uses linear interpolation with
\texttt{align\_corners=False}. These teacher computations are offline.

A causal trunk predicts non-negative utilities for all $K$ layers:
\begin{equation}
h_t=f_\theta(z_{\le t}),\qquad
\hat\Delta^o_{t,k}=\operatorname{softplus}(g_o(h_t)_k),
\quad o\in\{\mathrm{ac},\mathrm{sem}\},
\tag{8}
\end{equation}
without target-bitrate input. Layer 1 is mandatory; its predicted contribution
is identical across candidate depths and can therefore be omitted from the
allocation objective without changing depth selection.

Main acoustic-only and causal experiments use raw
$\hat\Delta^{\mathrm{ac}}$. Semantic-aware allocation is speech-only and
utterance-level, with each branch normalized separately:
\begin{equation}
s_o=\max\!\left(
\frac{1}{T(K-1)}
\sum_{t=1}^{T}\sum_{k=2}^{K}\hat\Delta^o_{t,k},10^{-8}\right),
\qquad
\bar\Delta^o_{t,k}=
\begin{cases}
0, & k=1,\\
\hat\Delta^o_{t,k}/s_o, & k=2,\ldots,K.
\end{cases}
\tag{9}
\end{equation}
Joint utility is
\begin{equation}
\Delta^{\mathrm{joint}}_{t,k}
=\bar\Delta^{\mathrm{ac}}_{t,k}
+\beta\bar\Delta^{\mathrm{sem}}_{t,k},\qquad \beta=0.25.
\tag{10}
\end{equation}
In semantic-weight sweeps, $\beta=0$ denotes the normalized acoustic reference,
not the main raw-acoustic system. Equation~(9) uses complete-utterance
statistics and is never used in causal streaming. For the selected utility,
\begin{equation}
U_t(d)=\sum_{k=2}^{d}u_{t,k},\qquad U_t(1)=0.
\tag{11}
\end{equation}

\subsection{Causal Exact-Budget Allocation}
\label{sec:online_allocator}

At frame $t$, candidate depth $d$ receives
\begin{equation}
S_t(d)=U_t(d)-\lambda_t c_t(d)
-\rho\,\mathbf{1}[d\neq d_{t-1}],
\tag{12}
\end{equation}
where $c_t(d)=10d$ is the code-index cost used as a rate-control surrogate in
the evaluated online configuration. The target per-frame cost is
$c=1000R_{\mathrm{target}}/f_{\mathrm{frame}}$, where
$R_{\mathrm{target}}$ is in kbps. Evaluation supplies this rate from the
matched fixed-depth serialized reference. The dual update is
\begin{equation}
\lambda_{t+1}
=\left[\lambda_t+\eta
\frac{c_t(d_t)-c}{\max(c,1)}\right]_+ .
\tag{13}
\end{equation}
A token bucket, dwell constraint, transition penalty, and hysteresis control
short-term rate and switching. No approximate depth-side cost is added to
$c_t(d)$ because the exact guard already accounts for depth-map RLE.

Exact serializer-consistent accounting uses
\begin{equation}
R_{\mathrm{actual}}
=H
+8\left\lceil\frac{R_{\mathrm{code}}}{8}\right\rceil
+8\left\lceil\frac{R^{\mathrm{RLE}}_{\mathrm{depth}}}{8}\right\rceil
+8\left\lceil\frac{R^{\mathrm{RLE}}_{\mathrm{expert}}}{8}\right\rceil ,
\tag{14}
\end{equation}
where $H$ is the fixed container-header cost and
$R_{\mathrm{code}}(d_{1:t})=10\sum_{i=1}^{t}d_i$ for 1024-entry codebooks.
The three variable payloads are independently byte aligned. Matched fixed and
dynamic prefixes have identical header and expert-map costs. Consequently,
the guard compares exact code-plus-depth payload sizes without knowing the
next expert assignment, while still enforcing the complete serialized budget.

Let $\tilde R$ denote this code-plus-depth payload and
$\tilde B_t=\tilde R^{\mathrm{fixed}}(D,t)$. Before commitment,
\begin{equation}
\mathcal F_t=
\left\{d\in\{1,\ldots,K\}:
\tilde R(d_{1:t-1},d)\le\tilde B_t,\;
\tilde R(d_{1:t-1},d,1)\le\tilde B_{t+1}
\right\}.
\tag{15}
\end{equation}
The second condition checks payload feasibility for a hypothetical next frame
at minimum depth $d=1$. Bucket affordability and dwell further restrict
admissible actions; minimum depth is exempt from the bucket mask. The
highest-scoring admissible depth is selected, with hysteresis retaining
$d_{t-1}$ when it remains admissible and the score improvement is below its
margin. Dwell is released when retaining the previous depth would violate
exact feasibility. Exact feasibility therefore takes precedence over all soft
switching controls.

Utterance-level experiments instead use full-sequence Lagrange--Viterbi
allocation with raw acoustic or normalized joint utility, followed by exact
serialized-budget checking. This offline branch is not used for causal
streaming.

\subsection{Prefix Representation and Dynamic Bitstream}
\label{sec:dynamic_representation}

A committed depth defines the active RVQ prefix. With fixed expert
assignments, prefix early exit is numerically equivalent to full RVQ evaluation
followed by masking. Our reported real-backbone evaluations use full RVQ
evaluation and the selected prefix mask; runtime results therefore do not rely
on computational early exit.

The dynamic container stores the fixed header, RLE-compressed depth and
block-shared expert maps, and packed active indices. The frozen decoder
reconstructs
\begin{equation}
\hat x=G\!\left(\{\tilde v_t(d_t)\}_{t=1}^{T}\right).
\tag{16}
\end{equation}
All metadata costs are included in $R_{\mathrm{actual}}$. Different budgets
select framewise prefixes of the same frozen hierarchy rather than separate
rate-specific tokenizers.

Depth decisions are causal per latent frame. Shared-expert quantization
buffers 75 frames (500 ms), using the majority Top-1 router assignment within
each block. Encoder, controller, allocator, and decoder states persist across
input chunks, so allocation does not reset at chunk boundaries. Reported
streaming timings include online exact-budget accounting but exclude final
byte packing and container parsing.

\subsection{Learning Objective and Training Protocol}
\label{sec:learning}

The codec and HuBERT teacher remain frozen. Utilities are trained directly
against non-negative marginal labels:
\begin{equation}
\mathcal L_o=
\frac{1}{|\Omega|}
\sum_{(t,k)\in\Omega}
\operatorname{SmoothL1}
\left(\hat\Delta^o_{t,k},\Delta^o_{t,k}\right),
\quad o\in\{\mathrm{ac},\mathrm{sem}\},
\tag{17}
\end{equation}
where $\Omega$ contains non-padded positions and all $K$ outputs.

We first train the acoustic controller. For semantic-head training, we
initialize from this checkpoint, freeze the shared causal trunk and acoustic
and speech-probability heads, and optimize only the semantic head. Its
regression loss is weighted by $20$, distinct from the allocation weight
$\beta=0.25$. HuBERT is absent from token extraction and deployment.
Architecture, data sizes, allocator hyperparameters, and streaming settings
are reported in the Appendix and experimental setup.

\section{Experiments}
\label{sec:experiments}

\subsection{Experimental Setup}
\label{sec:experimental_setup}

\paragraph{Codec and controllers.}
We use the frozen 48-kHz causal UniStream codec
\citep{zhao2026unistream}, with eight RVQ layers, 1024 entries per codebook,
and a latent rate of 150 Hz. We use reversible Top-1 shared-expert
quantization with one expert assignment per 500-ms block. The codec remains
frozen and only the utility controllers are trained. We evaluate target depths
$D\in\{2,3,4,6\}$, corresponding to approximately 3.1, 4.6, 6.1, and
9.1 kbps under fixed-depth coding. All UniAdapt rates are measured from the
actual serialized container, including active indices, depth RLE, expert RLE,
the header, and byte padding.

The speech acoustic controller uses 5,000 LibriSpeech train-clean-100
utterances \citep{panayotov2015librispeech}. The three-domain controller uses
3,000 examples, equally drawn from LibriSpeech, MTG-Jamendo
\citep{bogdanov2019mtg}, and FSD50K \citep{fonseca2022fsd50k}. Semantic
targets are generated from 1,000 LibriSpeech utterances using frozen HuBERT.
The three-domain offline evaluation uses 200 LibriSpeech test-clean
utterances, 50 MUSDB18-HQ~\citep{rafii2019musdb18hq} test examples,
and 50 FSD50K test examples.
Online allocation, online ablations, and streaming use the speech 5K
controller; the same 200 LibriSpeech test-clean utterances are evaluated at
all four online rates.

\paragraph{Evaluation and statistics.}
Reconstruction is evaluated using multi-resolution Log-STFT distance; speech
additionally uses PESQ and STOI \citep{rix2001pesq,taal2011stoi}.
Downstream probes use five independent seeds (2040--2044), with paired 95\%
Student-$t$ intervals over seeds ($df=4$). These intervals characterize
training-seed variability on a fixed evaluation set. Frozen external ASR uses
10,000 paired bootstrap resamples clustered by speaker, while external-codec
comparisons use 10,000 utterance-paired bootstrap resamples. MUSHRA uses the
listener as the paired statistical unit. Unless stated otherwise, confidence
intervals are pointwise and unadjusted for multiple comparisons. Full dataset
splits, probe architectures, secondary metrics, and implementation details are
reported in the Appendix.

\subsection{Utterance-Level Exact-Budget Allocation}
\label{sec:offline_results}

\begin{table}[t]
\caption{
Relative Log-STFT change of learned dynamic allocation versus fixed depth
under matched serialized-bit budgets. Results use utterance-level offline
allocation with the three-domain 3K controller. Percentages are computed from
the ratio of mean distortions; lower is better.
}
\label{tab:offline_main}
\centering
\vspace{0.4em}
\small

\begin{tabular*}{0.78\linewidth}{
@{\extracolsep{\fill}}
lcccc
@{}
}
\specialrule{0.8pt}{0pt}{0pt}
\textbf{Domain}
& \textbf{$D=2$}
& \textbf{$D=3$}
& \textbf{$D=4$}
& \textbf{$D=6$} \\
\specialrule{0.4pt}{2pt}{2pt}

Speech
& $-3.82\%$
& $-4.39\%$
& $-4.22\%$
& $-3.34\%$ \\

Music
& $-1.43\%$
& $-1.16\%$
& $-1.64\%$
& $-1.77\%$ \\

Environment
& $-1.53\%$
& $-1.07\%$
& $-1.61\%$
& $-1.34\%$ \\

\specialrule{0.8pt}{2pt}{0pt}
\end{tabular*}
\end{table}

Table~\ref{tab:offline_main} shows improvements at all four rates in all three
domains: 3.34--4.39\% for speech, 1.16--1.77\% for music, and
1.07--1.61\% for environmental audio. Every dynamic condition uses no more
serialized bits than its matched fixed-depth reference. On speech, the
three-domain controller captures 95.4--97.1\% of the latent-utility oracle
improvement under this offline allocation protocol. This gain-capture statistic
is computed from mean distortions and should not be conflated with the
utterance-level causal statistic below.

\subsection{Causal Online Allocation and Streaming}
\label{sec:online_results}

\begin{table}[t]
\caption{
Causal online allocation relative to matched fixed-depth coding. Percentages
are averages of per-utterance relative changes. Negative rate and Log-STFT
changes indicate lower bitrate and lower distortion, respectively.
}
\label{tab:online_main}
\centering
\small
\setlength{\tabcolsep}{4.5pt}
\begin{tabular}{@{}lrrrr@{}}
\toprule
\textbf{Target}
& \textbf{Rate $\Delta$}
& \textbf{Log-STFT $\Delta$}
& \textbf{Switches/s}
& \textbf{Violations} \\
\midrule
$D=2$ & $-3.56\%$ & $-0.70\%$ & 12.45 & 0/200 \\
$D=3$ & $-2.73\%$ & $+0.55\%$ & 17.14 & 0/200 \\
$D=4$ & $-2.86\%$ & $-0.91\%$ &  9.10 & 0/200 \\
$D=6$ & $-3.13\%$ & $-1.01\%$ &  8.11 & 0/200 \\
\bottomrule
\end{tabular}
\end{table}

Under causal commitment, UniAdapt improves Log-STFT at $D=2,4,6$, while
$D=3$ incurs a 0.55\% increase. All four operating points use fewer serialized
bits, with zero prefix-budget violations across 800 utterance--rate
evaluations.

At $D=4$, learned online allocation without dwell improves Log-STFT by
1.21\%, compared with 4.67\% for the offline latent oracle; the mean
utterance-level oracle-gain capture is 21.05\%. This differs from the
95.4--97.1\% offline statistic both in allocation constraints and aggregation:
the offline value is derived from mean distortions, whereas the online value
averages utterance-level gain ratios.

\paragraph{Streaming.}
On an NVIDIA A100, with the sequential allocator running on CPU, we evaluate
50 utterances at four rates and 100-, 200-, and 500-ms chunks. Chunked and
full causal execution produce identical transmitted indices, expert maps, and
depth maps; decoded waveforms agree within numerical tolerance. Mean
waveform-path RTF is 0.491--0.655, so all evaluated configurations run faster
than real time. The timed path includes online exact-budget accounting but
excludes final byte packing and container parsing. For 500-ms chunks,
estimated startup latency---500-ms expert buffering plus measured first-output
computation---is 735.9--745.2 ms, and side information occupies
1.89--6.25\% of the serialized container.

\paragraph{Ablations.}
The online controls trade reconstruction quality, switching, and budget
utilization. The ``no smoothing'' configuration jointly disables the
transition penalty, hysteresis, and dwell controls; it is not a separate
temporal filtering operation. Switching then increases from 9.10 to
37.05 transitions/s. Removing dwell improves the $D=4$ Log-STFT point
estimate from $-0.91\%$ to $-1.21\%$, showing that dwell primarily controls
trajectory stability rather than directly improving reconstruction.

With dwell reduced to one frame, learned acoustic utility improves Log-STFT by
1.21\%, while latent oracle utility improves it by 1.39\%. Three random
permutations of the flattened frame--layer utility array instead degrade
Log-STFT by 1.75--1.80\%, while uniform utility recovers fixed-depth
behavior. These results support the importance of assigning predicted
utilities to their corresponding frame--layer refinements rather than
introducing depth variation alone.

We further compare exact-prefix enforcement with an approximate-side-cost
configuration that disables the exact guard, restores a 3-bit/frame depth
charge, and uses dwell${}=1$. This alternative produces no final serialized
budget violations in the evaluated samples but underuses the matched budget by
6.50--12.81\%. The per-prefix hard guarantee follows from the feasibility
constraint in Sec.~\ref{sec:online_allocator}; this finite-sample comparison
instead supports improved budget utilization with exact accounting. Complete
ablation results are provided in the Appendix.

\subsection{Subjective and Objective-Dependent Evaluation}
\label{sec:semantic_subjective}

\begin{table}[t]
\caption{
Three-domain MUSHRA results using reconstructions produced by utterance-level
offline allocation. Confidence intervals use the listener as the paired
statistical unit.
}
\label{tab:mushra_main}
\centering
\small
\setlength{\tabcolsep}{5pt}
\begin{tabular}{@{}lrrrr@{}}
\toprule
\textbf{Domain}
& \textbf{Fixed}
& \textbf{Dynamic}
& \textbf{$\Delta$}
& \textbf{95\% CI} \\
\midrule
Speech
& 72.88 & 76.40 & $+3.52$ & $[+2.12,+4.92]$ \\
Music
& 65.33 & 65.84 & $+0.51$ & $[-0.96,+1.97]$ \\
Environment
& 66.06 & 64.85 & $-1.22$ & $[-2.71,+0.28]$ \\
Overall
& 68.09 & 69.03 & $+0.94$ & $[-0.20,+2.07]$ \\
\bottomrule
\end{tabular}
\end{table}

We conduct a MUSHRA-style study following ITU-R BS.1534
\citep{itu2015bs1534} using reconstructions produced by utterance-level
offline allocation rather than the causal streaming allocator. Twenty
listeners each complete 30 trials (10 speech, 10 music, and 10 environmental),
yielding 2,400 ratings; all satisfy the predefined hidden-reference quality
criterion. Each trial contains a hidden reference, fixed $D=4$, dynamic
$D=4$, and a 3.5-kHz low-pass anchor. Speech improves by 3.52 points with a
confidence interval excluding zero. Music, environmental audio, and overall
comparisons have intervals containing zero.

\begin{table}[t]
\caption{
Semantic-aware allocation relative to the corresponding normalized
acoustic-only reference at $\beta=0$. Results use development subsets;
oracle and learned evaluations use separately sampled 50-utterance subsets.
Negative semantic change is better, whereas positive Log-STFT change denotes
an acoustic cost.
}
\label{tab:semantic_main}
\centering
\small
\setlength{\tabcolsep}{7pt}
\begin{tabular}{@{}lrrr@{}}
\toprule
\textbf{Utility}
& \textbf{Depth}
& \textbf{Semantic $\Delta$}
& \textbf{Log-STFT $\Delta$} \\
\midrule
Oracle  & 2 & $-9.38\%$  & $+0.16\%$ \\
Oracle  & 3 & $-10.25\%$ & $+0.46\%$ \\
Oracle  & 4 & $-16.71\%$ & $+0.71\%$ \\
Learned & 2 & $-2.94\%$  & $+0.47\%$ \\
Learned & 3 & $-5.62\%$  & $+0.48\%$ \\
Learned & 4 & $-13.47\%$ & $+0.61\%$ \\
\bottomrule
\end{tabular}
\end{table}

On 20 LibriSpeech dev-clean utterances, signed HuBERT refinement gains used
for this diagnostic have an overall frame--layer correlation of 0.4205 with
the non-negative acoustic marginal gains. At
$D=4$, learned semantic-aware allocation reduces HuBERT distortion by
13.47\% at a 0.61\% Log-STFT cost. The semantic predictor has pooled
frame--layer Pearson correlation 0.733; this should not be interpreted as a
per-layer correlation. These experiments are speech-only and utterance-level;
causal streaming uses acoustic utility only.

\subsection{Scalable Frozen-Token Representations}
\label{sec:representation_results}

\begin{table}[t]
\caption{
Representative frozen-token results at $D=4$. Values are five-seed means.
Dynamic denotes acoustic utterance-level allocation; MTG uses the
artist-disjoint Top-50 downstream protocol. All rates include depth and expert
metadata.
}
\label{tab:downstream_main}
\centering
\small
\setlength{\tabcolsep}{5pt}
\begin{tabular}{@{}llrrr@{}}
\toprule
\textbf{Task}
& \textbf{Metric}
& \textbf{Fixed}
& \textbf{Dynamic}
& \textbf{Rate $\Delta$} \\
\midrule
CTC
& CER $\downarrow$
& 0.5993 & 0.5924 & $-0.57\%$ \\
Speaker
& EER $\downarrow$
& 0.1782 & 0.1616 & $-0.57\%$ \\
FSD50K
& macro-mAP $\uparrow$
& 0.0681 & 0.0697 & $-1.38\%$ \\
MTG
& macro-mAP $\uparrow$
& 0.1006 & 0.1007 & $-0.26\%$ \\
Clotho
& mean recall $\uparrow$
& 0.0230 & 0.0239 & $-0.60\%$ \\
\bottomrule
\end{tabular}
\end{table}

The shared cross-rate CTC probe is trained on dynamic multi-rate tokens and
yields lower CER at all four rates, with 5/5 seed wins and paired intervals
excluding zero. Because its greedy WER remains near one, we treat CER as a
lightweight cross-rate representation diagnostic rather than a replacement for
external ASR.

The speaker probe is likewise trained on dynamic multi-rate tokens. Dynamic
tokens significantly reduce EER at $D=3,4,6$, whereas $D=2$ is not
statistically conclusive; the improvement does not extend uniformly to the
secondary minDCF metric. For FSD50K, macro-mAP improves in all five seeds by
$+0.001633$, with 95\% CI $[+0.001204,+0.002062]$, while using
1.381\% fewer serialized bits.

For artist-disjoint MTG-Jamendo and
Clotho~\citep{drossos2020clotho}, the primary metric point estimates
are positive at lower realized bitrates, but their paired intervals include
zero. We therefore report no statistically significant difference on these
metrics; the comparisons do not establish equivalence or non-inferiority.
For MTG, artist-disjoint refers specifically to the downstream probe split.

\paragraph{Frozen external ASR.}
At $D=4$, fixed, acoustic-dynamic, and semantic-joint reconstructions obtain
WERs of 2.762\%, 2.630\%, and 2.594\%, respectively, using frozen wav2vec2
\citep{baevski2020wav2vec}. The joint condition gives a 6.11\% relative WER
reduction in the point estimate, but the 10,000-resample speaker-cluster
bootstrap interval reaches zero. We therefore report this as a strong trend
rather than a statistically conclusive recognition improvement.

\subsection{External Codec Positioning}
\label{sec:external_codec}

On all 2,620 LibriSpeech test-clean utterances, the official 24-kHz EnCodec
checkpoint~\citep{defossez2023encodec} obtains PESQ/STOI 2.747/0.938 at
6.113 kbps, compared with 1.913/0.871 at 6.076 kbps for UniAdapt fixed and
2.043/0.889 at 6.043 kbps for UniAdapt dynamic. Dynamic allocation
significantly improves the same frozen UniAdapt backbone at a slightly lower
rate, while EnCodec retains higher absolute reconstruction quality. Because
the systems differ in architecture, sample rate, training objective, and
quantization design, this comparison is contextual rather than
architecture-controlled.
\section{Conclusion}
\label{sec:conclusion}

We introduced UniAdapt, which treats audio tokens as a budgeted
representation resource by learning frame- and layer-wise marginal utilities
and allocating RVQ capacity causally under the true serialized-bit budget.
The same frozen hierarchy supports multiple rates and acoustic or semantic
objectives without retraining the codec.

Across speech, music, and environmental audio, UniAdapt improves
utterance-level reconstruction under no-larger serialized budgets. In the
causal setting, it improves three of four speech rates with zero budget
violations and runs faster than real time. Human evaluation shows a significant speech-quality improvement, while
frozen-token experiments show selective downstream gains and support
cross-rate use.


\section*{AI Use Statement}

We used generative AI tools to assist with code debugging, experiment
bookkeeping, figure preparation, and language drafting and editing.
The authors designed the research questions, methods, experimental protocols,
and statistical analyses. The authors manually reviewed all AI-assisted text,
verified reported numerical results against the corresponding experiment
files, tested AI-assisted code for correctness, and checked citations against
the original sources. Generative AI tools did not generate experimental
measurements or listener ratings. The authors take responsibility for the final content of the paper
and the reported results.

\section*{Ethics Statement}

This work uses publicly available speech, music, and environmental audio
datasets under their respective licenses. The subjective evaluation involved
human listeners who were informed about the listening task and participated
voluntarily. No personally sensitive information was collected beyond the
responses required for the listening test.

Neural audio coding can support efficient communication and audio
representation, but reconstructed speech and speaker representations may also
raise privacy concerns. We use speaker verification only as a representation
benchmark and do not perform demographic or sensitive-attribute inference.
\section*{Reproducibility Statement}

The paper and Appendix specify the frozen codec and reversible operating mode,
utility targets and controller architecture, training procedure, exact
serialized-bit accounting, online allocation configuration, operating rates,
streaming protocol, evaluation settings, random seeds, and statistical
procedures. The Appendix additionally reports implementation details,
correctness and equivalence checks, controller training configurations, and
the checkpoints used for the reported experiments.

\bibliography{iclr2027_conference}
\bibliographystyle{iclr2027_conference}


\newpage
\appendix

\section{Implementation and Reproducibility Details}
\label{app:implementation}

This section provides implementation details omitted from the main paper.
We distinguish between utterance-level acoustic or semantic-aware allocation
and causal acoustic depth allocation. Joint utility uses complete-utterance
normalization, whereas the causal acoustic controller uses raw predicted
utilities. Causality of the online allocation policy is conditional on the
supplied budget reference and target rate; shared-expert quantization
additionally requires block buffering.

\subsection{Frozen Codec and Reversible Operating Mode}
\label{app:codec}

All experiments use a frozen UniStream backbone
\citep{zhao2026unistream}. The codec operates at $48$~kHz and produces
$150$ latent frames per second. Its quantizer contains $K=8$ RVQ layers:
one shared first layer and seven expert-specific acoustic layers. Each
codebook contains $1024$ entries, so each active RVQ index requires
$10$ bits. The encoder, quantizer input projection, codebooks, routing
modules, and decoder remain frozen.

The encoder produces $z_t\in\mathbb{R}^{C}$, and the quantizer input
projection initializes the residual:
\begin{equation}
    v_t=P_{\mathrm{in}}(z_t)\in\mathbb{R}^{C_q},
    \qquad
    r_{t,0}=v_t,
    \qquad
    r_{t,k}=r_{t,k-1}-q_{t,k}.
    \label{eq:app_rvq_residual}
\end{equation}
The controller consumes encoder latents $z_t$, while acoustic utility
is measured in the projected quantization space.

For expert-specific layers, we derive frame-level Top-1 assignments from
the first acoustic router and take their majority within each block of
$75$ latent frames ($500$~ms). The selected expert is shared across the
acoustic RVQ layers in that block. Expert assignments are explicitly
serialized and are not decision variables of the depth allocator.
The backbone has four acoustic experts; the utility controller uses a
single depth-action expert dimension because it predicts depth utility
under this externally determined expert map.

Serialized experiments use the reversible Top-1/shared-expert mode.
The legacy Top-2 forward path mixes expert outputs using continuous
routing weights, but its stored discrete representation does not contain
sufficient information to reproduce that mixture exactly. We therefore
do not use that path for UniAdapt bitstreams.

The main fixed reference depths are
\[
    D\in\{2,3,4,6\},
\]
with realized fixed-container rates of approximately
$3.1$, $4.6$, $6.1$, and $9.1$~kbps. Dynamic conditions are compared
against the corresponding serialized fixed-depth reference, including
its side information, rather than against the nominal
$150\times D\times10$ index rate.

\subsection{Utility Targets and Controller}
\label{app:controller}

\paragraph{Acoustic targets.}
For frame $t$ and layer $k$, the target is the non-negative reduction
in channel-normalized residual energy:
\begin{equation}
    \Delta^{\mathrm{ac}}_{t,k}
    =
    \left[
        \frac{\|r_{t,k-1}\|_2^2}{C_q}
        -
        \frac{\|r_{t,k}\|_2^2}{C_q}
    \right]_+,
    \qquad k=1,\ldots,K.
    \label{eq:app_acoustic_target}
\end{equation}
The implementation predicts all $K$ layers. Because layer $1$ is
mandatory, its contribution is identical across candidate depths and
can be omitted from the allocation objective without changing depth
selection.

\paragraph{Semantic targets.}
For speech, we use the frozen
\texttt{facebook/hubert-base-ls960} teacher
\citep{hsu2021hubert}, taking its layer-9 hidden states. Waveforms are
resampled from $48$ to $16$~kHz before teacher inference. Let
$\hat{x}^{(k)}$ denote reconstruction at uniform RVQ depth $k$.
Original and reconstructed waveforms are restricted to their common
sample length, and teacher outputs are restricted to their common
frame length $J$.

On the HuBERT time grid, semantic distortion is cosine distance:
\begin{equation}
    D^{\mathrm{sem}}_{j,k}
    =
    1-\operatorname{cos}
    \left(
        \Phi_j(x),\Phi_j(\hat{x}^{(k)})
    \right),
    \qquad j=1,\ldots,J.
    \label{eq:app_semantic_distance}
\end{equation}
We first difference adjacent-depth distortions and clip negative gains,
then interpolate to the codec grid:
\begin{equation}
    \Delta^{\mathrm{sem}}_{t,k}
    =
    \mathcal I_{J\rightarrow T}
    \left(
        \left[
            D^{\mathrm{sem}}_{\cdot,k-1}
            -
            D^{\mathrm{sem}}_{\cdot,k}
        \right]_+
    \right)_t,
    \quad k=2,\ldots,K,
    \qquad
    \Delta^{\mathrm{sem}}_{t,1}=0.
    \label{eq:app_semantic_target}
\end{equation}
Interpolation uses \texttt{mode="linear"} and
\texttt{align\_corners=False}. This order of differencing, clipping,
and interpolation matches label generation. HuBERT is used offline
for semantic targets and semantic-distortion evaluation; it is absent
from token extraction and deployed inference.

\paragraph{Prediction and training loss.}
The controller directly predicts non-negative marginal utilities:
\begin{equation}
    h_t=f_\theta(z_{\le t}),
    \qquad
    \hat\Delta^o_{t,k}
    =
    \operatorname{softplus}\!\left(g_o(h_t)_k\right),
    \quad o\in\{\mathrm{ac},\mathrm{sem}\}.
    \label{eq:app_utility_prediction}
\end{equation}
Targets are used in their original utility domain. No logarithmic target
transformation or inverse transformation is applied. Training uses
masked Smooth-L1 regression:
\begin{equation}
    \mathcal L_o
    =
    \frac{1}{|\Omega|}
    \sum_{(t,k)\in\Omega}
    \operatorname{SmoothL1}
    \left(
        \hat\Delta^o_{t,k},\Delta^o_{t,k}
    \right),
    \label{eq:app_utility_loss}
\end{equation}
where $\Omega$ contains non-padded positions and all $K$ layer outputs.

\paragraph{Causal architecture.}
The shared trunk contains three left-padded temporal convolutions
with kernel sizes $5$, $3$, and $3$ and channel dimensions
$256\rightarrow128\rightarrow128\rightarrow64$.
Each convolution is followed by channel LayerNorm applied independently
at each frame and a GELU activation. Separate $1\times1$ convolutional
heads predict acoustic and semantic utilities, followed by softplus.

The three convolutions require $4$, $2$, and $2$ previous latent frames,
respectively. Exact chunked controller inference therefore retains
eight raw latent frames, equivalent to $53.33$~ms at $150$~Hz.
The target bitrate is never supplied to the predictor.
A speech-probability head is present in the implementation, but its loss
weight is zero in the reported configurations and its output is not used
for acoustic or joint allocation.

\paragraph{Optimization.}
Controllers use AdamW with learning rate $3\times10^{-4}$ and gradient
norm clipping at $1.0$. The speech acoustic controller uses $5{,}000$
LibriSpeech train-clean-100 utterances. The balanced three-domain
controller uses $1{,}000$ examples each from LibriSpeech, MTG-Jamendo,
and the official FSD50K training split; its training records are
randomly cropped to at most $450$ latent frames.

For semantic-head training, we initialize from the speech acoustic
checkpoint and freeze the shared trunk, acoustic head, and
speech-probability head. Only the semantic head is optimized, using
$1{,}000$ speech examples and a semantic regression weight of $20$.
This training weight is distinct from the allocation weight
$\beta=0.25$.

\subsection{Exact Joint-Utility Normalization}
\label{app:joint_normalization}

Semantic-aware joint experiments normalize acoustic and semantic
utilities separately before combining them. Let $U^{(m)}_{t,\ell}$
denote a predicted or oracle marginal utility for objective
$m\in\{\mathrm{ac},\mathrm{sem}\}$. For each utterance and branch,
we compute one scalar over all time frames and optional layers:
\begin{equation}
    s_m
    =
    \max\left(
        \frac{1}{T(K-1)}
        \sum_{t=1}^{T}
        \sum_{\ell=2}^{K}
        U^{(m)}_{t,\ell},
        10^{-8}
    \right).
    \label{eq:app_joint_scale}
\end{equation}
Normalization is
\begin{equation}
    \widetilde U^{(m)}_{t,\ell}
    =
    \begin{cases}
        0, & \ell=1,\\[2mm]
        U^{(m)}_{t,\ell}/s_m, & \ell=2,\ldots,K.
    \end{cases}
    \label{eq:app_joint_normalization}
\end{equation}
The two branches use distinct denominators
$s_{\mathrm{ac}}$ and $s_{\mathrm{sem}}$. Statistics are computed
over all optional layers, not only those selected by the allocation.

The combined utility is
\begin{equation}
    U^{\mathrm{joint}}_{t,\ell}
    =
    \widetilde U^{(\mathrm{ac})}_{t,\ell}
    +
    \beta\,\widetilde U^{(\mathrm{sem})}_{t,\ell},
    \qquad \beta=0.25.
    \label{eq:app_joint_utility}
\end{equation}
There is no subsequent renormalization or division by $1+\beta$.
When neither denominator is clamped, each normalized optional-layer
branch has mean one and the joint mean is $1.25$.
Setting the first-layer utility to zero does not remove its code:
layer $1$ remains mandatory.

\paragraph{Causality and baseline distinction.}
Complete-utterance normalization is used only for offline joint
token extraction and semantic oracle/learned-allocation experiments.
In semantic-weight sweeps, $\beta=0$ denotes the normalized acoustic
reference. The main acoustic-only token experiments and causal
streaming experiments instead consume raw predicted acoustic utilities.
They neither compute utterance-global utility normalization nor
combine semantic utility. No speech-probability or bitrate gate is
applied in the reported joint experiments.

\subsection{Online Allocator Configuration}
\label{app:online_config}

The causal experiments use the speech-specific acoustic controller.
The formal online configuration was selected on development data
and fixed for test evaluation:
\begin{equation}
\begin{aligned}
    \eta &= 0.002, &
    \rho &= 1.0,\\
    h &= 0.20, &
    m_{\mathrm{dwell}} &= 5,\\
    T_{\mathrm{bucket}} &= 0.5~\mathrm{s}, &
    b_{\mathrm{init}} &= 0.
\end{aligned}
\end{equation}
Here $\eta$ is the dual step size, $\rho$ the transition penalty,
$h$ the hysteresis margin, and $m_{\mathrm{dwell}}$ the dwell-counter
threshold in latent frames.

\begin{table}[t]
\caption{
Initial dual prices used for the four online operating points.
}
\label{tab:app_initial_dual}
\centering
\small
\setlength{\tabcolsep}{8pt}
\begin{tabular}{@{}lcccc@{}}
\toprule
\textbf{Target depth} & \textbf{2} & \textbf{3} & \textbf{4} & \textbf{6} \\
\midrule
$\lambda_{\mathrm{init}}$
& 0.20 & 0.10 & 0.08 & 0.02 \\
\bottomrule
\end{tabular}
\end{table}

Let $R_{\mathrm{target}}$ be the supplied target rate in kbps and
$f_{\mathrm{frame}}=150$~Hz. In evaluation,
$R_{\mathrm{target}}$ is the actual container bitrate of the matched
fixed-depth reference for that utterance. The rate-control quantities are
\begin{equation}
    c=\frac{1000R_{\mathrm{target}}}{f_{\mathrm{frame}}},
    \qquad
    c_t(d)=10d,
    \qquad
    b_{\max}=1000R_{\mathrm{target}}T_{\mathrm{bucket}}.
    \label{eq:app_rate_control_cost}
\end{equation}
The code-index cost $c_t(d)$ is used as a surrogate for total
serialized cost. Additional estimated depth-side and expert-side
charges are zero in the final online configuration; actual side-map
costs remain included in the serialized-budget comparison.

For the utility $U_t(d)$ defined in the main text, the candidate score is
\begin{equation}
    S_t(d)
    =
    U_t(d)-\lambda_t c_t(d)
    -\rho\,\mathbf{1}[d\neq d_{t-1}].
    \label{eq:app_online_score}
\end{equation}
Bucket affordability requires $c_t(d)\le b_t+c$, with minimum depth
$d=1$ exempt from this bucket mask. Exact feasibility is enforced
separately and is never bypassed.

The dwell counter is initialized to $m_{\mathrm{dwell}}$, reset to
zero after a depth change, and incremented when depth is retained.
While the counter is below its threshold, the previous depth is
retained if it remains exactly feasible. Otherwise the dwell lock is
released. After applying admissibility masks, hysteresis retains the
previous depth when it remains admissible and the score improvement
is below $h$.

After choosing $d_t$, rate-control state is updated by
\begin{equation}
\begin{aligned}
    b_{t+1}
    &=
    \operatorname{clip}
    \left(b_t+c-c_t(d_t),\,0,\,b_{\max}\right),\\
    \lambda_{t+1}
    &=
    \left[
        \lambda_t+
        \eta\frac{c_t(d_t)-c}{\max(c,1)}
    \right]_+ .
\end{aligned}
\label{eq:app_online_updates}
\end{equation}
Initializing the bucket empty avoids granting initial rate credit.
The exact prefix guard, rather than the bucket alone, provides the
hard serialized-budget guarantee.

\subsection{Exact Serialized-Bit Accounting}
\label{app:bit_accounting}

The container consists of a fixed header, a depth RLE payload, an
expert RLE payload, and packed active indices. The header occupies
$H=256$ bits. Each variable payload is independently byte aligned.

For $1024$-entry codebooks,
\begin{equation}
    R_{\mathrm{code}}(d_{1:t})
    =
    10\sum_{i=1}^{t}d_i .
\end{equation}
A map run stores its value followed by an Elias-gamma-coded positive
run length. For the eight-layer depth map, each depth value occupies
four bits; for the four-expert map, each expert ID occupies two bits.
Writing $\mathcal R_d$ and $\mathcal R_e$ for their runs,
\begin{equation}
\begin{aligned}
    R^{\mathrm{RLE}}_{\mathrm{depth}}
    &=
    \sum_{r\in\mathcal R_d}
    \left(4+2\lfloor\log_2\ell_r\rfloor+1\right),\\
    R^{\mathrm{RLE}}_{\mathrm{expert}}
    &=
    \sum_{r\in\mathcal R_e}
    \left(2+2\lfloor\log_2\ell_r\rfloor+1\right).
\end{aligned}
\label{eq:app_rle_bits}
\end{equation}
Consecutive expert blocks selecting the same expert form one run.

The exact size obtained by serializing a sequence prefix is
\begin{equation}
\begin{aligned}
    R_{\mathrm{actual}}(d_{1:t},e_{1:t})
    = H
    &+8\left\lceil R_{\mathrm{code}}/8\right\rceil\\
    &+8\left\lceil R^{\mathrm{RLE}}_{\mathrm{depth}}/8\right\rceil\\
    &+8\left\lceil R^{\mathrm{RLE}}_{\mathrm{expert}}/8\right\rceil .
\end{aligned}
\label{eq:app_exact_bits}
\end{equation}
This is a sequence-prefix size definition, not a claim that each
prefix is separately packetized and transmitted.

Matched fixed and dynamic prefixes use the same header format and
expert map. Their header and expert-map costs are therefore identical.
Define the exact code-plus-depth payload
\begin{equation}
    \tilde R(d_{1:t})
    =
    8\left\lceil R_{\mathrm{code}}(d_{1:t})/8\right\rceil
    +
    8\left\lceil
        R^{\mathrm{RLE}}_{\mathrm{depth}}(d_{1:t})/8
    \right\rceil,
\end{equation}
and let $\tilde B_t=\tilde R(D,\ldots,D)$ for a length-$t$ fixed-depth
reference. The guard admits
\begin{equation}
\mathcal F_t=
\left\{
d\in\{1,\ldots,K\}:
\begin{aligned}
    \tilde R(d_{1:t-1},d)&\le\tilde B_t,\\
    \tilde R(d_{1:t-1},d,1)&\le\tilde B_{t+1}
\end{aligned}
\right\}.
\label{eq:app_guard}
\end{equation}
The second condition checks feasibility of a hypothetical next frame
at minimum depth. Because common side costs appear on both sides,
the guard does not need the next expert assignment to perform this
comparison. All header and expert-map bits nevertheless remain
included in the reported total bitrate.

The guard maintains incremental serializer-consistent state: the
active-code count, completed depth-run bits, current depth, and
current run length. It does not materialize every candidate bitstream
at every frame. The implementation was additionally stress-tested
using $200$ random allocation trajectories per target depth
$D\in\{2,3,4,6\}$, each extending to $5{,}000$ frames.
No state without a feasible action was observed in these checks.

\subsection{Streaming State and Chunked Inference}
\label{app:streaming_impl}

Encoder latents feed the cached causal utility controller and online
allocator. Buffered latents also undergo full shared-expert RVQ
evaluation. The committed depths select a prefix mask, and the selected
latent vectors are passed to the cached waveform decoder.
With fixed expert assignments, computational prefix early exit is an
equivalent execution option; reported real-backbone runtime experiments
use full RVQ evaluation followed by masking.

The encoder has receptive field $2254$ waveform samples and stride
$320$. Its cached implementation retains $2560$ stride-aligned waveform
samples ($53.33$~ms).

\begin{table}[t]
\caption{
Input chunk sizes and corresponding latent-frame counts used for
streaming evaluation.
}
\label{tab:app_chunk_frames}
\centering
\small
\setlength{\tabcolsep}{9pt}
\begin{tabular}{@{}lccc@{}}
\toprule
\textbf{Input chunk}
& \textbf{100 ms}
& \textbf{200 ms}
& \textbf{500 ms} \\
\midrule
Latent frames
& 15 & 30 & 75 \\
\bottomrule
\end{tabular}
\end{table}

Non-final input chunks are stride aligned. The controller retains eight
raw latent frames. The allocator carries its dual price, bucket state,
previous depth, and dwell counter across chunks, together with the
incremental exact-budget state. Its generic action-expert field is not
an optimization of the backbone expert map. The decoder retains eleven
latent frames of context ($73.33$~ms). Shared-expert processing buffers
$75$ frames and flushes any remaining partial block at the end of the
utterance.

The final runtime configuration places the sequential allocator on CPU
and the neural codec and controller on the GPU, avoiding repeated
per-frame GPU synchronization. Chunked execution is compared against
full causal execution under the same configuration.

Reported streaming timings include encoding, utility prediction, online
allocation and exact-budget accounting, full RVQ evaluation, prefix
masking, and waveform decoding. Final byte packing and container parsing
are excluded. Estimated startup latency is defined as the $500$-ms
expert-block buffer plus measured first-output computation.

\subsection{Correctness and Equivalence Checks}
\label{app:correctness}

We use explicit checks for chunked computation, depth feasibility, and
dynamic serialization.

For the cached encoder, five test utterances evaluated with
$100$, $200$, and $500$~ms chunks produce exactly the same latent
values as full encoding in the tested configuration, with zero maximum
absolute difference. Chunked utility prediction agrees with full
causal prediction within numerical tolerance, and chunked allocation
produces the same depth maps as processing the causal sequence in one
call.

Decoder checks cover five utterances, depths $2$, $4$, and $6$, and
latent chunk lengths $15$, $30$, and $75$. The reported maximum
waveform absolute difference is approximately
$2.94\times10^{-6}$, with relative $\ell_2$ difference approximately
$5.49\times10^{-7}$.

End-to-end checks compare active code indices, block-shared expert maps,
depth maps, serialized-bit counts, and decoded waveforms. Discrete
quantities match exactly; waveforms agree within numerical tolerance.
The formal streaming evaluation covers $50$ utterances, four rates,
and three chunk sizes, giving $600$ conditions with no observed
budget violations. These empirical checks complement the
serializer-consistent feasibility rule; they are not a proof based
solely on finite samples.

\subsection{Training Sets, Checkpoints, and Random Seeds}
\label{app:assets}

The principal controller training configurations are summarized in
Table~\ref{tab:app_controller_training}.

\begin{table}[t]
\caption{
Utility-controller training configurations. Speech acoustic and semantic
labels are generated from excerpts of at most three seconds; the
three-domain training loader randomly crops records to at most $450$
latent frames. Semantic training initializes from the speech acoustic
checkpoint and updates only the semantic head.
}
\label{tab:app_controller_training}
\centering
\small
\setlength{\tabcolsep}{5pt}
\begin{tabular}{@{}lrrr@{}}
\toprule
\textbf{Controller}
& \textbf{Examples}
& \textbf{Batch size}
& \textbf{Steps} \\
\midrule
Speech acoustic
& $5{,}000$ & 32 & $10{,}000$ \\
Three-domain acoustic
& $3{,}000$ & 16 & $10{,}000$ \\
Speech semantic head
& $1{,}000$ & 32 & $5{,}000$ \\
\bottomrule
\end{tabular}
\end{table}

Speech acoustic label sampling uses seed $2030$, semantic label
sampling uses seed $2038$, and balanced three-domain subset sampling
uses seed $2040$. These are data-sampling seeds, distinct from the
downstream probe training seeds $2040$--$2044$.

The formal online experiments use the speech acoustic controller at
step $10{,}000$. Semantic-aware experiments use the semantic-head
checkpoint at step $5{,}000$, and three-domain reconstruction uses the
balanced acoustic controller at step $10{,}000$. All conditions share
the frozen UniStream codec checkpoint at step $200{,}000$.

\section{Additional Allocation and Ablation Results}
  \label{app:allocation}

  This section provides additional results underlying
  Sec.~\ref{sec:online_results}. Unless otherwise stated, experiments use
  the same $200$ LibriSpeech test-clean utterances, target depth $D=4$,
  the frozen speech-specific 5K acoustic controller, and $500$-ms input
  chunks. Codec parameters and expert assignments remain fixed across
  allocation conditions.

  For condition $a$, let $L_i^a$ and $R_i^a$ denote the Log-STFT
  distortion and serialized bitrate of utterance $i$. Percentage changes
  in this section are averages of per-utterance relative changes:
  \begin{equation}
  \begin{aligned}
      \delta_L^a
      &=
      \frac{100}{N}\sum_{i=1}^{N}
      \left(\frac{L_i^a}{L_i^{\mathrm{fixed}}}-1\right),\\
      \delta_R^a
      &=
      \frac{100}{N}\sum_{i=1}^{N}
      \left(\frac{R_i^a}{R_i^{\mathrm{fixed}}}-1\right).
  \end{aligned}
  \label{eq:app_ablation_relative_changes}
  \end{equation}
  Negative values indicate lower distortion or bitrate. Switching rates
  are averaged over utterances. The violation column counts utterances
  whose final serialized size exceeds the matched fixed-depth budget.
  Conditions using the exact guard additionally enforce the per-prefix
  constraint defined in Sec.~\ref{sec:online_allocator}; zero final
  violations alone do not establish prefix feasibility for unguarded
  conditions.

  \subsection{Online Allocator Component Ablations}
  \label{app:allocator_components}

  The \emph{Full} configuration is the locked online configuration used
  in the main experiments: transition penalty $\rho=1.0$, hysteresis
  margin $h=0.20$, dwell-counter threshold $m=5$, dual step
  $\eta=0.002$, bucket duration $0.5$~s, an initially empty bucket,
  and the exact prefix-feasibility guard. At $D=4$, the initial dual
  price is $0.08$.

  The ``no smoothing'' condition jointly sets $\rho=0$, $h=0$, and
  $m=1$; it does not remove a separate temporal filter. ``No dwell''
  denotes the experimental setting $m=1$ throughout this section.
  ``Penalty only,'' ``hysteresis only,'' and ``dwell only'' retain
  only the named switching control, while leaving dual rate control
  and exact-budget enforcement active. ``No dual update'' sets
  $\eta=0$ while retaining the initial dual price. ``Redundant side
  cost'' adds a $3$-bit/frame estimated depth charge while retaining
  the exact guard.

  \begin{table}[t]
  \caption{
Online component ablations at $D=4$ on the same $200$ test utterances.
Changes are relative to matched fixed-depth coding and use
Eq.~(\ref{eq:app_ablation_relative_changes}). All configurations in
this table retain the exact prefix guard.
}
  \label{tab:app_allocator_ablation}
  \centering
  \small
  \setlength{\tabcolsep}{4pt}
  \begin{tabular}{@{}lrrrr@{}}
  \toprule
  \textbf{Configuration}
  & \textbf{Rate $\Delta$}
  & \textbf{Log-STFT $\Delta$}
  & \textbf{Switches/s}
  & \textbf{Violations}\\
  \midrule
  Full, $m=5$
  & $-2.860\%$ & $-0.906\%$ & 9.10 & $0/200$\\
  No smoothing
  & $-0.411\%$ & $-0.674\%$ & 37.05 & $0/200$\\
  No transition penalty
  & $-1.286\%$ & $-0.445\%$ & 18.93 & $0/200$\\
  No hysteresis
  & $-2.653\%$ & $-0.843\%$ & 10.84 & $0/200$\\
  No dwell, $m=1$
  & $-3.060\%$ & $-1.206\%$ & 9.70 & $0/200$\\
  Penalty only
  & $-2.665\%$ & $-1.212\%$ & 11.43 & $0/200$\\
  Hysteresis only
  & $-1.164\%$ & $-1.142\%$ & 20.12 & $0/200$\\
  Dwell only
  & $-0.987\%$ & $-0.231\%$ & 23.66 & $0/200$\\
  No dual update
  & $-17.213\%$ & $-0.060\%$ & 7.16 & $0/200$\\
  Redundant side cost
  & $-9.526\%$ & $+1.085\%$ & 9.69 & $0/200$\\
  \bottomrule
  \end{tabular}
  \end{table}

  Jointly disabling the switching controls raises transitions from
  $9.10$ to $37.05$/s. Removing only the transition penalty raises
  switching to $18.93$/s and reduces the reconstruction improvement.
  These results show that switching controls regulate the temporal
  complexity of the depth map and its associated side-information cost.

  Dwell exhibits a quality--stability trade-off. Reducing its threshold
  from five to one improves the Log-STFT change from $-0.906\%$ to
  $-1.206\%$, while increasing switching from $9.10$ to $9.70$/s.
  Penalty-only and hysteresis-only configurations also have better
  reconstruction point estimates than Full, but higher switching rates.
  Thus, the controls do not each contribute a monotonic quality gain.

  Without dual updates, the allocator uses $17.21\%$ less bitrate than
  the matched reference and recovers almost no reconstruction gain.
  This demonstrates the importance of adapting the dual price under
  the tested initialization and configuration. Adding an estimated
  depth-side charge on top of the exact guard likewise leads to
  substantial budget under-utilization and worse reconstruction.
  The final online configuration therefore uses zero estimated
  depth-side charge while retaining exact depth-RLE accounting.

  \paragraph{Development-set dwell comparison.}
  We additionally compare the two dwell settings on the full
  $500$-utterance development set.

  \begin{table}[t]
  \caption{
  Development-set comparison at $D=4$. Both conditions use the same
  controller, exact guard, transition penalty, and hysteresis; only the
  dwell threshold changes. Percentages average per-utterance changes.
  }
  \label{tab:app_dwell_dev}
  \centering
  \small
  \setlength{\tabcolsep}{5pt}
  \begin{tabular}{@{}lrrrr@{}}
  \toprule
  \textbf{Configuration}
  & \textbf{Rate $\Delta$}
  & \textbf{Log-STFT $\Delta$}
  & \textbf{Switches/s}
  & \textbf{Violations}\\
  \midrule
  Full, $m=5$
  & $-2.7554\%$ & $-1.0680\%$ & 9.562 & $0/500$\\
  No dwell, $m=1$
  & $-2.5693\%$ & $-1.3806\%$ & 9.971 & $0/500$\\
  \bottomrule
  \end{tabular}
  \end{table}

  The development results show the same qualitative trade-off:
  $m=1$ improves reconstruction but increases switching. The paired
  change in relative Log-STFT distortion is $-0.3126$ percentage
  points, while switching increases by $0.4089$/s. We retain the
  locked $m=5$ configuration for the main streaming results and
  report $m=1$ separately.

  \subsection{Utility-Information Baselines}
  \label{app:utility_ablation}

  We next hold the $D=4$ allocator fixed with $m=1$ and vary its
  utility input. Learned utility is the acoustic controller output.
  For each shuffled control, the frame--layer utility array is
  flattened, randomly permuted, and reshaped. This preserves the
  multiset of utility values within each utterance but disrupts
  both temporal and layer correspondence. Uniform utility assigns
  the utterance-wide mean predicted utility to every frame--layer
  entry.

  The oracle input uses ground-truth non-negative marginal reductions
  in projected latent residual energy. It is a privileged diagnostic
  requiring full RVQ computations, not a deployable predictor or a
  guaranteed optimum for waveform distortion.

  \begin{table}[t]
  \caption{
  Utility-information controls at $D=4$ with the same online allocator,
  exact prefix guard, and dwell threshold $m=1$. Shuffled controls
  permute flattened frame--layer entries. Uniform and oracle inputs
  use privileged full-utterance information; causality here describes
  the allocator's decision recurrence, not every utility-generation
  procedure.
  }
  \label{tab:app_utility_ablation}
  \centering
  \small
  \setlength{\tabcolsep}{4pt}
  \begin{tabular}{@{}lrrrr@{}}
  \toprule
  \textbf{Utility}
  & \textbf{Rate $\Delta$}
  & \textbf{Log-STFT $\Delta$}
  & \textbf{Switches/s}
  & \textbf{Violations}\\
  \midrule
  Learned acoustic
  & $-3.0596\%$ & $-1.2062\%$ & 9.697 & $0/200$\\
  Shuffled, seed 2040
  & $-0.0092\%$ & $+1.7947\%$ & 26.282 & $0/200$\\
  Shuffled, seed 2041
  & $-0.0062\%$ & $+1.7467\%$ & 26.824 & $0/200$\\
  Shuffled, seed 2042
  & $-0.0117\%$ & $+1.7978\%$ & 26.851 & $0/200$\\
  Uniform
  & $0.0000\%$ & $0.0000\%$ & 0.000 & $0/200$\\
  Oracle latent utility
  & $-3.1503\%$ & $-1.3892\%$ & 12.556 & $0/200$\\
  \bottomrule
  \end{tabular}
  \end{table}

  All three shuffled controls degrade Log-STFT by approximately
  $1.75$--$1.80\%$, whereas learned utility improves it by $1.21\%$.
  This supports the importance of assigning utilities to their
  corresponding frame--layer refinements rather than introducing
  arbitrary depth variation. Because shuffling also changes layer
  correspondence, this comparison does not isolate temporal alignment
  alone.

  Uniform utility recovers fixed-depth behavior in this configuration.
  Replacing predicted utility with ground-truth latent utility
  improves the relative Log-STFT change by $0.1830$ percentage points.
  The gain is modest under the same allocator, although the realized
  bitrates and switching rates also differ. This comparison measures
  the effect of replacing the utility signal within the tested
  allocation procedure, rather than an isolated causal decomposition
  of all sources of reconstruction error.

  \subsection{Exact versus Approximate Budget Enforcement}
  \label{app:exact_guard_ablation}

  We compare exact-prefix enforcement with an approximate-side-cost
  configuration. The alternative disables the exact guard, restores
  a depth-side charge of $3$ bits/frame, and uses $m=1$. Initial
  dual prices remain $0.20$, $0.10$, $0.08$, and $0.02$ for
  $D=2,3,4,6$, respectively; the remaining rate-control parameters
  follow the online configuration.

  This is a comparison between budget mechanisms, not an ablation
  that changes only guard availability. In particular, the score
  and bucket costs change when the approximate side charge is
  restored.

  \begin{table}[t]
  \caption{
  Approximate-side-cost allocation without the exact guard. The
  allocator uses a $3$-bit/frame depth-side charge and $m=1$.
  Violations count final serialized-budget overruns; their absence
  does not establish that every intermediate prefix is feasible.
  }
  \label{tab:app_no_exact_guard}
  \centering
  \small
  \setlength{\tabcolsep}{5pt}
  \begin{tabular}{@{}crrrr@{}}
  \toprule
  \textbf{Target}
  & \textbf{Rate $\Delta$}
  & \textbf{Log-STFT $\Delta$}
  & \textbf{Switches/s}
  & \textbf{Violations}\\
  \midrule
  $D=2$
  & $-12.81\%$ & $+0.02\%$ & 16.36 & $0/200$\\
  $D=3$
  & $-9.03\%$ & $+0.75\%$ & 19.60 & $0/200$\\
  $D=4$
  & $-8.80\%$ & $+0.04\%$ & 13.67 & $0/200$\\
  $D=6$
  & $-6.50\%$ & $-0.12\%$ & 12.33 & $0/200$\\
  \bottomrule
  \end{tabular}
  \end{table}

  No final serialized-budget overrun occurs across these
  $800$ utterance--rate evaluations. Nevertheless, the alternative
  uses $6.50$--$12.81\%$ less bitrate than the matched reference
  and provides no positive reconstruction point-estimate gain at
  $D=2,3,4$.

  The hard per-prefix guarantee of the guarded method follows from
  its feasibility constraint: every committed action must satisfy
  the exact current-prefix budget and next-frame minimum-depth
  payload condition. The finite-sample comparison instead provides
  empirical evidence of better budget utilization under exact
  accounting than under the tested approximate-side-cost mechanism.
  It does not show that removing the guard alone necessarily causes
  final-budget violations.

  \subsection{Online versus Offline Oracle}
  \label{app:online_offline_gap}

  We compare learned and ground-truth latent utilities within the
  same online allocator and then contrast them with full-utterance
  latent-oracle allocation. All rows use the same $200$ test
  utterances and matched $D=4$ fixed-depth budgets. The online
  conditions use $m=1$ and exact-prefix enforcement. The offline
  reference uses full-sequence Lagrange--Viterbi allocation with
  \texttt{switch\_bits}$=50$ and an exact final-container constraint.

  \begin{table}[t]
  \caption{
  Learned and oracle-utility allocation at $D=4$. Percentages average
  per-utterance distortion changes relative to fixed depth. The
  online oracle row supplies privileged latent utilities to the
  online allocator; the offline row uses full-utterance allocation.
  The latent oracle is a proxy reference, not a certified optimum
  for waveform Log-STFT distortion.
  }
  \label{tab:app_online_offline_gap}
  \centering
  \small
  \setlength{\tabcolsep}{5pt}
  \begin{tabular}{@{}lrr@{}}
  \toprule
  \textbf{Condition}
  & \textbf{Log-STFT $\Delta$}
  & \textbf{Switches/s}\\
  \midrule
  Learned online, $m=1$
  & $-1.2062\%$ & 9.697\\
  Oracle utility, online allocator
  & $-1.3892\%$ & 12.556\\
  Offline latent oracle
  & $-4.6748\%$ & 5.688\\
  \bottomrule
  \end{tabular}
  \end{table}

  Replacing learned utility with ground-truth latent utility within
  the same online allocator changes the relative Log-STFT improvement
  from $1.2062\%$ to $1.3892\%$. The offline latent oracle achieves
  $4.6748\%$, leaving a considerably larger difference between
  allocation procedures.

  There are two distinct ways to express the learned-online versus
  offline-oracle difference. Relative to the common fixed-depth
  baseline, the reported percentage changes differ by
  \[
      (-1.2062)-(-4.6748)=3.4686
      \quad\text{percentage points}.
  \]
  Alternatively, averaging the distortion ratio relative to the
  offline oracle gives
  \begin{equation}
      G_{\mathrm{rel}}
      =
      \frac{100}{N}
      \sum_{i=1}^{N}
      \left(
          \frac{L_i^{\mathrm{online}}}
               {L_i^{\mathrm{offline\text{-}oracle}}}
          -1
      \right)
      =
      3.6639\%.
      \label{eq:app_online_relative_gap}
  \end{equation}
  The latter is a relative percentage, not a percentage-point
  difference between the two changes from fixed depth.

  The per-utterance oracle-gain capture is
  \begin{equation}
      G_{\mathrm{capture}}
      =
      \frac{100}{|\mathcal J|}
      \sum_{i\in\mathcal J}
      \frac{
          L_i^{\mathrm{fixed}}-L_i^{\mathrm{online}}
      }{
          L_i^{\mathrm{fixed}}
          -L_i^{\mathrm{offline\text{-}oracle}}
      }
      =
      21.05\%,
      \label{eq:app_online_gain_capture}
  \end{equation}
  where
  $\mathcal J=\{i:
  L_i^{\mathrm{fixed}}>L_i^{\mathrm{offline\text{-}oracle}}\}$.
  This is a mean of utterance-level ratios, rather than a ratio of
  mean improvements.

  The main three-domain offline evaluation reports
  $95.4$--$97.1\%$ speech oracle-gain capture computed from mean
  distortions. That result uses a different controller and evaluation
  protocol, as well as different allocation constraints and
  aggregation. It should not be interpreted as the gain capture of
  the causal streaming system.

  These comparisons indicate that improving utility prediction alone,
  while retaining the tested online allocator, recovers only a small
  part of the observed difference to the offline reference. However,
  the online and offline procedures also differ in lookahead, prefix
  constraints, switching treatment, and optimization strategy.
  The experiments therefore do not isolate an unavoidable cost of
  causality or establish an upper bound on achievable online quality.
  They identify online allocation as an important remaining source
  of improvement alongside utility modeling.

\section{Streaming and Bitstream Analysis}
  \label{app:streaming}

  This section reports the streaming measurements and correctness checks
  underlying Sec.~\ref{sec:online_results}. All experiments use the frozen
  48-kHz codec and speech-specific acoustic utility controller. Encoder
  latents feed the causal controller and online allocator, while buffered
  latents undergo full shared-expert RVQ evaluation. The committed depths
  select a prefix mask, and the selected latent vectors are reconstructed
  by the cached decoder.

  The timed waveform-to-waveform path includes encoding, utility prediction,
  online allocation and exact-budget accounting, full RVQ evaluation,
  prefix masking, and waveform decoding. It excludes model loading,
  audio-file loading and preprocessing, final byte packing, and container
  parsing. The reported runtime results therefore do not rely on
  computational prefix early exit or measure network transmission latency.

  The input chunk sizes are $100$, $200$, and $500$~ms, corresponding to
  $15$, $30$, and $75$ latent frames at $150$~Hz. Streaming states persist
  across chunk boundaries. Shared-expert quantization buffers $75$ frames
  and applies one expert assignment across the acoustic RVQ layers of
  each block; any remaining partial block is processed at the end of
  the utterance.

  \subsection{Chunked--Full Equivalence}
  \label{app:streaming_equivalence}

  We compare chunked execution with full-sequence causal computation under
  the same controller, allocator configuration, and supplied budget
  reference. Full-sequence causal allocation processes the utility sequence
  in one call while retaining the same sequential decision rule; it is
  distinct from the offline Lagrange--Viterbi allocator.

  The encoder has a receptive field of $2254$ waveform samples and total
  stride $320$. The cached implementation retains $2560$ stride-aligned
  waveform samples ($53.33$~ms). Across five test utterances and all three
  chunk sizes, the chunked and full encoder outputs have identical shapes
  and zero measured latent difference.

  The utility controller caches eight raw latent frames ($53.33$~ms).
  Its chunked outputs agree with full causal outputs within numerical
  tolerance. The allocator preserves its dual price, token-bucket state,
  previous depth, dwell counter, and incremental exact-budget state across
  chunks. Its generic action-expert field does not optimize the backbone
  expert map. Chunked and one-shot causal allocation produce identical
  framewise depth trajectories in the evaluated conditions.

  The cached decoder retains $11$ latent frames ($73.33$~ms).
  Checks cover five utterances, depths $D\in\{2,4,6\}$, and latent
  chunk lengths $15$, $30$, and $75$, giving $45$ conditions.
  The observed maximum waveform absolute difference is approximately
  $2.94\times10^{-6}$, and the maximum relative $\ell_2$ difference
  is approximately $5.49\times10^{-7}$.

  End-to-end comparisons additionally check:
  \begin{itemize}
      \item encoder latents and predicted acoustic utilities;
      \item quantizer layer vectors and RVQ code indices;
      \item block-shared expert maps and framewise depth maps;
      \item total serialized-bit counts; and
      \item reconstructed waveforms.
  \end{itemize}
  Code indices, expert maps, depth maps, and serialized sizes match
  exactly; continuous tensors agree within numerical tolerance.

  An initial evaluation covers $20$ utterances, four rates, and three
  chunk sizes, giving $240$ conditions. The final evaluation covers
  $50$ utterances under the same rate--chunk combinations, giving
  $600$ conditions. All pass the corresponding equivalence and
  serialized-budget checks.

  \subsection{End-to-End Real-Time Performance}
  \label{app:streaming_rtf}

  The final runtime configuration uses an NVIDIA A100 GPU for the neural
  codec and controller and CPU execution for the sequential allocator.
  Rate--chunk configurations are measured serially to avoid contention
  from concurrent allocator processes.

  For utterance $i$, waveform-path real-time factor is
  \begin{equation}
      \mathrm{RTF}_i
      =
      \frac{t_i^{\mathrm{processing}}}
           {T_i^{\mathrm{audio}}},
      \label{eq:app_streaming_rtf}
  \end{equation}
  where $t_i^{\mathrm{processing}}$ is elapsed time for the complete
  chunk-processing loop and $T_i^{\mathrm{audio}}$ is input duration.
  The waveform is loaded before timing, and chunks are processed
  without waiting for real-time input arrival. GPU synchronization
  is used in the timing procedure.

  \begin{table}[t]
  \caption{
  Waveform-streaming performance over $50$ utterances.
  RTF entries are mean $\pm$ across-utterance standard deviation.
  Timing includes online exact-budget accounting, full RVQ evaluation,
  and prefix masking, but excludes final byte packing and parsing.
  Allocation statistics are averaged over utterances and are identical
  across the three chunk sizes for each target.
  }
  \label{tab:app_streaming_rtf}
  \centering
  \small
  \setlength{\tabcolsep}{3.8pt}
  \begin{tabular}{@{}crrrrrr@{}}
  \toprule
  &
  \multicolumn{3}{c}{\textbf{Waveform-path RTF}}
  &
  \multicolumn{3}{c}{\textbf{Allocation Statistics}}\\
  \cmidrule(lr){2-4}
  \cmidrule(lr){5-7}
  \textbf{Target}
  & \textbf{100 ms}
  & \textbf{200 ms}
  & \textbf{500 ms}
  & \textbf{Rate $\Delta$}
  & \textbf{Avg.\ depth}
  & \textbf{Switches/s}\\
  \midrule
  $D=2$
  & $0.655\pm0.040$
  & $0.575\pm0.013$
  & $0.491\pm0.015$
  & $-2.327\%$ & 1.878 & 12.45\\
  $D=3$
  & $0.649\pm0.013$
  & $0.574\pm0.013$
  & $0.492\pm0.015$
  & $-2.203\%$ & 2.843 & 16.40\\
  $D=4$
  & $0.636\pm0.014$
  & $0.543\pm0.014$
  & $0.510\pm0.015$
  & $-2.699\%$ & 3.831 & 9.21\\
  $D=6$
  & $0.643\pm0.014$
  & $0.543\pm0.013$
  & $0.515\pm0.015$
  & $-2.976\%$ & 5.760 & 9.19\\
  \bottomrule
  \end{tabular}
  \end{table}

  All twelve rate--chunk configurations have mean RTF below one.
  The ranges are $0.636$--$0.655$ for $100$-ms chunks,
  $0.543$--$0.575$ for $200$-ms chunks, and
  $0.491$--$0.515$ for $500$-ms chunks. Larger chunks require
  fewer calls and less repeated processing of cached context per
  unit of audio, consistent with their lower measured RTF.

  Rate changes are averages of per-utterance relative differences
  against the matched fixed-depth serialized containers.
  These statistics use the $50$-utterance streaming subset and
  therefore differ from the $200$-utterance online-allocation
  statistics in the main paper. Every evaluated streaming condition
  satisfies its matched serialized budget.

  \subsection{Estimated Startup Latency and Output Processing Intervals}
  \label{app:streaming_latency}

  We separately profile first-output computation and subsequent output
  completion intervals using $500$-ms chunks. Let $c_{i,b}$ denote
  elapsed processing time from the start of the chunk loop until
  output block $b$ is completed for utterance $i$. Estimated startup
  latency is
  \begin{equation}
      \hat{\tau}^{\mathrm{startup}}_i
      =
      500~\mathrm{ms}+c_{i,1},
      \label{eq:app_startup_latency}
  \end{equation}
  with $c_{i,1}$ expressed in milliseconds. This adds the required
  expert-block accumulation time to measured first-output computation.
  It is not a measurement of model-loading latency or a complete
  capture--network--playback system.

  For an utterance with $M_i>1$ output blocks, we also compute
  \begin{equation}
      \bar{\tau}^{\mathrm{processing}}_i
      =
      \frac{1}{M_i-1}
      \sum_{b=2}^{M_i}
      \left(c_{i,b}-c_{i,b-1}\right).
      \label{eq:app_output_interval}
  \end{equation}
  This statistic measures intervals between output completions while
  processing preloaded audio without input pacing. It can include the
  final partial block and should not be interpreted as a network
  packet interval or the output cadence of a real-time audio source.

  \begin{table}[t]
  \caption{
Separate $500$-ms latency-profile runs over $50$ utterances.
Startup is estimated using Eq.~(\ref{eq:app_startup_latency}).
The processing interval is defined by
Eq.~(\ref{eq:app_output_interval}); its mean and standard deviation
are computed across utterance-level means. All $\pm$ entries are
standard deviations, not confidence intervals. Overhead fractions
and rates are means across utterances.
}
  \label{tab:app_streaming_latency}
  \centering
  \small
  \setlength{\tabcolsep}{3.8pt}
  \begin{tabular}{@{}crrrrr@{}}
  \toprule
  \textbf{Target}
  & \textbf{Est.\ startup}
  & \textbf{Processing interval}
  & \textbf{Overhead}
  & \textbf{Overhead kbps}
  & \textbf{RTF}\\
  \midrule
  $D=2$
  & $744.8\pm6.4$ ms
  & $243.5\pm3.6$ ms
  & $6.25\%$ & 0.187 & $0.514\pm0.015$\\
  $D=3$
  & $744.1\pm6.4$ ms
  & $244.4\pm3.3$ ms
  & $4.71\%$ & 0.211 & $0.515\pm0.015$\\
  $D=4$
  & $745.2\pm7.0$ ms
  & $243.1\pm3.5$ ms
  & $2.79\%$ & 0.165 & $0.513\pm0.015$\\
  $D=6$
  & $735.9\pm7.4$ ms
  & $234.9\pm10.4$ ms
  & $1.89\%$ & 0.166 & $0.496\pm0.025$\\
  \bottomrule
  \end{tabular}
  \end{table}

  Estimated startup ranges from $735.9$ to $745.2$~ms, while mean
  output processing intervals range from $234.9$ to $244.4$~ms.
  These quantify different aspects of execution: startup includes
  expert-block accumulation, whereas the processing interval
  excludes waiting for input arrival. The implementation achieves
  faster-than-real-time processing in these measurements while
  retaining substantial initial buffering.

  The RTF values in Table~\ref{tab:app_streaming_latency} come from
  separate profiling runs and need not equal the $500$-ms RTF
  estimates in Table~\ref{tab:app_streaming_rtf}.

  \subsection{Serialized Side-Information and Padding Overhead}
  \label{app:side_information}

  The dynamic container explicitly stores the depth trajectory and
  shared-expert map. Reported side-information overhead includes
  the header, byte-aligned depth and expert maps, and code-payload
  padding. For utterance $i$, let $p_i^{\mathrm{code}}$ be the
  number of padding bits added to the packed active indices.
  The measured overhead is
  \begin{equation}
  \begin{aligned}
      S_i = H
      &+8\left\lceil
          R^{\mathrm{RLE}}_{\mathrm{depth},i}/8
      \right\rceil\\
      &+8\left\lceil
          R^{\mathrm{RLE}}_{\mathrm{expert},i}/8
      \right\rceil
      +p_i^{\mathrm{code}}.
  \end{aligned}
  \label{eq:app_side_information}
  \end{equation}
  The reported fraction and bitrate are
  \begin{equation}
      \frac{100S_i}{R_{\mathrm{actual},i}}
      \quad\text{and}\quad
      \frac{S_i}{1000T_i^{\mathrm{audio}}},
      \label{eq:app_overhead_metrics}
  \end{equation}
  respectively, averaged across utterances. Thus, these values
  represent all bits beyond the raw active-code indices, not only
  the two side maps.

  For the $500$-ms profiles, overhead accounts for
  $1.89$--$6.25\%$ of the serialized container, corresponding to
  $0.165$--$0.211$~kbps. Its relative fraction is largest at the
  lowest evaluated rate, where fewer code-index bits are available
  to amortize the overhead.

  The $500$-ms expert block trades signaling cost against buffering.
  Sharing assignments reduces expert-map variation, and adjacent
  blocks with the same assignment can be represented by a single
  RLE run. The corresponding expert block must nevertheless be
  available before its quantized output is finalized.
  All reported rates include these costs; no packet-transport
  overhead is measured.

  \subsection{Runtime Placement and Component Profile}
  \label{app:streaming_profile}

  The allocator makes a small sequential decision at each latent
  frame. An initial implementation ran this loop on GPU and incurred
  repeated host--device synchronization. In a representative
  $D=4$, $500$-ms profile, total RTF was approximately $4.47$,
  with allocator RTF approximately $4.04$.

  The final implementation runs this control loop on CPU while
  keeping the encoder, utility controller, RVQ, and decoder on GPU.
  This preserves the allocation rule; the corresponding executions
  are checked against their full causal references.

  \begin{table}[t]
  \caption{
  Representative $D=4$, $500$-ms profile with CPU allocation.
  These are measurements for one profiling utterance, not the
  $50$-utterance means in Table~\ref{tab:app_streaming_rtf}.
  End-to-end RTF also includes processing-loop overhead not assigned
  to an individual component.
  }
  \label{tab:app_component_rtf}
  \centering
  \small
  \setlength{\tabcolsep}{8pt}
  \begin{tabular}{@{}lr@{}}
  \toprule
  \textbf{Component} & \textbf{RTF}\\
  \midrule
  Cached encoder & 0.0379\\
  Utility controller & 0.0071\\
  CPU allocator and exact guard & 0.0698\\
  Full shared-expert RVQ & 0.2756\\
  Cached decoder & 0.0992\\
  Other processing-loop overhead & 0.0120\\
  \midrule
  End-to-end waveform path & 0.5017\\
  \bottomrule
  \end{tabular}
  \end{table}

  CPU placement reduces total RTF in these representative profiles
  from approximately $4.47$ to $0.502$. Full RVQ evaluation is the
  largest measured component in the final profile.
  The difference between end-to-end time and the sum of instrumented
  stage times accounts for additional loop operations, including
  prefix masking and buffering. The component totals may differ
  slightly after rounding.

  These measurements characterize the evaluated implementation
  and hardware. They do not claim that CPU execution is universally
  preferable for all allocators or that the reported runtime includes
  a separately measured serialization pipeline.

  \subsection{Streaming Budget Correctness}
  \label{app:streaming_budget}

  The chunked implementation uses the same exact prefix-feasibility
  guard as the one-shot causal allocator. Before a depth is
  committed, it must satisfy the matched fixed-depth prefix budget
  and the next-frame minimum-depth payload condition.

  The guard incrementally tracks the exact byte-aligned code and
  depth-map payloads. Matched fixed and dynamic prefixes use
  identical header and expert-map costs, so the payload comparison
  enforces the complete serialized-budget inequality without
  requiring future expert assignments. Header, expert-map, and
  padding bits remain included when the final container size is
  computed.

  All
  \[
      50\ \text{utterances}
      \times4\ \text{target rates}
      \times3\ \text{chunk sizes}
      =600
  \]
  final streaming conditions satisfy the matched serialized budget.
  Their chunked and full causal executions also agree on code
  indices, expert assignments, depth maps, and serialized sizes.

  The empirical checks support chunked--full equivalence and
  consistency between guard accounting and the serializer on the
  evaluated data. The per-prefix budget guarantee follows from
  restricting committed actions to the exactly feasible set, rather
  than from the absence of violations in a finite test set.
  Here a serialized prefix means the size obtained by serializing
  the corresponding sequence prefix; these experiments do not
  measure independently packetized prefix transmission.

\section{Semantic and Objective-Dependent Analysis}
  \label{app:semantic}

  This section examines objective-dependent refinement value and evaluates
  semantic-aware allocation on speech. All joint-allocation results use the
  utterance-level normalization in Appendix~\ref{app:joint_normalization}.
  The codec remains frozen, and causal streaming uses acoustic utility only.

  \subsection{Marginal Gains Across RVQ Layers}
  \label{app:semantic_depth}

  We analyze 20 LibriSpeech dev-clean utterances using frozen HuBERT
  features \citep{hsu2021hubert}. Mean cosine distortion decreases from
  $0.3755$ at uniform depth 1 to $0.0483$ at depth 8, with diminishing
  average gains from additional layers.

  For this diagnostic, semantic gains retain their signs:
  \begin{equation}
      \delta^{\mathrm{sem}}_{j,k}
      =
      D^{\mathrm{sem}}_{j,k-1}-D^{\mathrm{sem}}_{j,k}.
      \label{eq:app_signed_semantic_gain}
  \end{equation}
  Unlike the non-negative training targets, these diagnostic gains are
  not clipped. Their means and positive-frame fractions are computed
  on the HuBERT grid and averaged over utterances. For correlation
  analysis, signed gains are linearly interpolated to the codec grid
  and compared with the non-negative acoustic marginal gains.

  \begin{table}[t]
  \caption{
  Optional-layer semantic gains on 20 development utterances.
  Gain means and positive-frame fractions use signed HuBERT-grid
  differences; Pearson correlations use aligned codec-grid entries.
  }
  \label{tab:app_semantic_layer_corr}
  \centering
  \small
  \setlength{\tabcolsep}{7pt}
  \begin{tabular}{@{}crrr@{}}
  \toprule
  \textbf{Layer}
  & \textbf{Mean gain}
  & \textbf{Positive frames}
  & \textbf{Acoustic correlation}\\
  \midrule
  2 & 0.182084 & 93.9\% &  0.2875\\
  3 & 0.077078 & 92.9\% &  0.1147\\
  4 & 0.033587 & 87.8\% &  0.0164\\
  5 & 0.015633 & 83.7\% & $-0.0200$\\
  6 & 0.008651 & 80.1\% & $-0.0102$\\
  7 & 0.006105 & 76.2\% & $-0.0231$\\
  8 & 0.004106 & 72.4\% & $-0.0364$\\
  \bottomrule
  \end{tabular}
  \end{table}

  Pooling optional layers and frames gives Pearson correlation
  $0.4205$, whereas within-layer correlations approach zero in deeper
  layers. The pooled statistic also reflects between-layer variation
  and should not be interpreted as uniformly strong frame-level
  agreement. These observations motivate modeling semantic value
  separately from acoustic residual reduction; they do not establish
  statistical independence or semantic disentanglement.

  \subsection{Oracle and Learned Semantic Allocation}
  \label{app:learned_semantic}

  Oracle allocation uses measured acoustic and semantic marginal
  utilities, while learned allocation uses controller predictions.
  Both normalize each objective separately and compare joint allocation
  with its corresponding normalized acoustic reference at $\beta=0$.
  All conditions satisfy the same matched fixed-depth budget ceiling;
  their realized serialized sizes need not be identical.

  Oracle and learned evaluations use separately sampled 50-utterance
  dev-clean subsets, with sampling seeds 2034 and 2039, respectively.
  Percentage changes are computed from mean distortions within each
  evaluation. The two subsets are not paired with each other.

  \begin{table}[t]
  \caption{
  Speech semantic--acoustic trade-offs under utterance-level allocation.
  Each row is relative to the corresponding $\beta=0$ reference.
  Negative semantic change is better; positive Log-STFT change is an
  acoustic cost. Oracle and learned rows use separate development
  subsets and are not a paired predictor comparison.
  }
  \label{tab:app_semantic_oracle}
  \label{tab:app_learned_semantic}
  \centering
  \small
  \setlength{\tabcolsep}{6pt}
  \begin{tabular}{@{}lcrrr@{}}
  \toprule
  \textbf{Utility}
  & \textbf{$\beta$}
  & \textbf{Depth}
  & \textbf{Semantic $\Delta$}
  & \textbf{Log-STFT $\Delta$}\\
  \midrule
  Oracle & 0.25 & 2 & $-9.38\%$  & $+0.16\%$\\
  Oracle & 0.25 & 3 & $-10.25\%$ & $+0.46\%$\\
  Oracle & 0.25 & 4 & $-16.71\%$ & $+0.71\%$\\
  Oracle & 0.50 & 2 & $-11.49\%$ & $+0.87\%$\\
  Oracle & 0.50 & 3 & $-14.34\%$ & $+1.35\%$\\
  Oracle & 0.50 & 4 & $-23.65\%$ & $+1.50\%$\\
  \midrule
  Learned & 0.25 & 2 & $-2.94\%$  & $+0.47\%$\\
  Learned & 0.25 & 3 & $-5.62\%$  & $+0.48\%$\\
  Learned & 0.25 & 4 & $-13.47\%$ & $+0.61\%$\\
  \bottomrule
  \end{tabular}
  \end{table}

  Increasing $\beta$ shifts oracle allocation toward lower HuBERT
  distortion at greater acoustic cost. We use $\beta=0.25$ for the
  reported learned joint condition as a moderate trade-off, without
  claiming it is universally optimal. The oracle is a privileged
  utility reference, not a certified optimum for reconstructed
  waveform quality.

  The semantic head is trained for 5,000 steps using 1,000 speech
  utterances, with the shared trunk and acoustic head frozen.
  On 80 held-out utility records, pooled prediction--target Pearson
  correlations are $0.9906$ for acoustic utility and $0.7329$ for
  semantic utility; the corresponding $R^2$ values are $0.9792$ and
  $0.5357$. These aggregate statistics include all layer outputs and
  do not imply comparable accuracy within each layer.

  Despite weaker semantic prediction, learned joint allocation at
  $D=4$ reduces mean HuBERT distortion from $0.125546$ to $0.108630$,
  a $13.47\%$ reduction, with a $0.61\%$ Log-STFT increase.
  Thus, objective-dependent allocation remains effective without
  modifying the frozen codec.

  \subsection{Frozen External ASR}
  \label{app:frozen_asr}

  Because HuBERT distortion also defines the semantic training target,
  we additionally evaluate reconstructions using a separate frozen
  ASR model, \texttt{facebook/wav2vec2-large-960h-lv60-self},
  with greedy CTC decoding \citep{baevski2020wav2vec}.
  Evaluation uses 500 LibriSpeech test-clean utterances from
  40 speakers. All reconstructed conditions use utterance-level
  allocation at $D=4$.

  \begin{table}[t]
  \caption{
  Frozen external ASR on 500 test-clean utterances.
  CER and WER are corpus-level error rates; bitrate is the mean
  serialized-container bitrate. Bold identifies the best point
  estimate among reconstructed conditions, not statistical significance.
  }
  \label{tab:app_frozen_asr}
  \centering
  \small
  \setlength{\tabcolsep}{7pt}
  \begin{tabular}{@{}lrrr@{}}
  \toprule
  \textbf{Condition}
  & \textbf{CER (\%)}
  & \textbf{WER (\%)}
  & \textbf{kbps}\\
  \midrule
  Original waveform
  & 0.5479 & 1.9180 & --\\
  Fixed $D=4$
  & 0.8447 & 2.7624 & 6.0823\\
  Acoustic dynamic $D=4$
  & 0.8516 & 2.6297 & 6.0476\\
  Semantic joint $D=4$
  & \textbf{0.8379} & \textbf{2.5935} & \textbf{6.0424}\\
  \bottomrule
  \end{tabular}
  \end{table}

  Joint allocation reduces the WER point estimate by $0.1689$
  percentage points, or $6.11\%$ relative to fixed depth, while using
  a lower bitrate. We assess uncertainty using 10,000 paired
  speaker-cluster bootstrap resamples, recomputing corpus error
  rates from the sampled speaker totals.

  The pointwise 95\% WER-difference intervals, in percentage points,
  are $[-0.3813,+0.1302]$ for dynamic minus fixed,
  $[-0.3439,0.0000]$ for joint minus fixed, and
  $[-0.2641,+0.1858]$ for joint minus dynamic.
  All include zero, so the ASR results support a favorable point
  estimate but not a statistically conclusive improvement.

  Direct HuBERT-distortion analyses use development subsets, whereas
  this external ASR evaluation uses held-out test utterances.
  Together, the results support a semantic--acoustic allocation
  trade-off, with more limited evidence of downstream recognition
  benefits. They do not establish semantic-aware causal streaming
  or generalization of the semantic head to non-speech audio.

\section{Full Frozen-Token Representation Results}
  \label{app:representation}

  We freeze the codec and utility controllers before downstream training.
  Probes consume active RVQ indices and their depth/expert metadata;
  inactive layers are masked. ``Dynamic'' denotes acoustic utterance-level
  allocation, while ``joint'' denotes speech-only semantic-aware
  utterance-level allocation. All rates include serialized side information.

  Experiments use five probe-training seeds, $2040$--$2044$.
  Condition summaries report mean $\pm$ standard deviation; paired
  differences report mean $\pm$ the half-width of a pointwise 95\%
  Student-$t$ interval with $df=4$. These intervals characterize seed
  variability on fixed evaluation data and are unadjusted for multiple
  comparisons. Differences are calculated before rounding.

  \subsection{Cross-Rate Character Recognition}
  \label{app:crossrate_ctc}

  For each seed, one shared CTC probe is trained on dynamic tokens at
  $D\in\{2,3,4,6\}$. The $5{,}000$ source utterances yield $20{,}000$
  utterance--rate pairs. The encoder uses active-layer masking,
  aggregation normalized by the square root of active depth,
  $4\times$ temporal subsampling, and a two-layer Transformer.
  Training runs for $20{,}000$ steps, with checkpoint selection based
  on mean development CER across the four dynamic conditions.

  Each probe is evaluated on fixed, dynamic, and joint tokens at all
  four rates on the same 500 test utterances, giving 60 evaluations
  across five seeds.

  \begin{table}[t]
  \caption{
  Cross-rate CTC results. CER and WER are fractions, not percentages;
  error rates are mean $\pm$ standard deviation across five seeds.
  Bitrates are means over test utterances and are shared across seeds.
  }
  \label{tab:app_crossrate_ctc_full}
  \centering
  \small
  \setlength{\tabcolsep}{4pt}
  \begin{tabular}{@{}clrrr@{}}
  \toprule
  \textbf{Depth} & \textbf{Condition}
  & \textbf{CER} & \textbf{WER} & \textbf{kbps}\\
  \midrule
  2 & Fixed
  & $0.614976\pm0.027256$ & $1.008878\pm0.006890$ & 3.080988\\
  & Dynamic
  & $0.610068\pm0.025802$ & $1.008154\pm0.006409$ & 3.052069\\
  & Joint
  & $0.607274\pm0.026936$ & $1.009530\pm0.006924$ & 3.045245\\
  \midrule
  3 & Fixed
  & $0.601594\pm0.029039$ & $1.009650\pm0.006415$ & 4.582023\\
  & Dynamic
  & $0.594388\pm0.026446$ & $1.009867\pm0.008351$ & 4.551451\\
  & Joint
  & $0.595302\pm0.026673$ & $1.010856\pm0.007094$ & 4.543908\\
  \midrule
  4 & Fixed
  & $0.599269\pm0.027084$ & $1.005959\pm0.006793$ & 6.082280\\
  & Dynamic
  & $0.592434\pm0.026855$ & $1.008999\pm0.009020$ & 6.047587\\
  & Joint
  & $0.592233\pm0.027896$ & $1.011508\pm0.007661$ & 6.042355\\
  \midrule
  6 & Fixed
  & $0.604717\pm0.027794$ & $1.008323\pm0.006420$ & 9.084434\\
  & Dynamic
  & $0.597712\pm0.026994$ & $1.008106\pm0.006189$ & 9.045284\\
  & Joint
  & $0.597580\pm0.026467$ & $1.007527\pm0.006790$ & 9.038676\\
  \bottomrule
  \end{tabular}
  \end{table}

  \begin{table}[t]
  \caption{
  Paired CER differences across five seeds.
  Each $\pm$ value is the half-width of a pointwise 95\% Student-$t$
  interval. All intervals exclude zero; both comparisons improve
  in all five seeds at every depth.
  }
  \label{tab:app_crossrate_ctc_diff}
  \centering
  \small
  \setlength{\tabcolsep}{9pt}
  \begin{tabular}{@{}crr@{}}
  \toprule
  \textbf{Depth}
  & \textbf{Dynamic $-$ Fixed}
  & \textbf{Joint $-$ Fixed}\\
  \midrule
  2 & $-0.004908\pm0.002745$ & $-0.007703\pm0.000525$\\
  3 & $-0.007205\pm0.003887$ & $-0.006292\pm0.003339$\\
  4 & $-0.006835\pm0.001601$ & $-0.007036\pm0.001498$\\
  6 & $-0.007004\pm0.003676$ & $-0.007137\pm0.002906$\\
  \bottomrule
  \end{tabular}
  \end{table}

  Dynamic and joint tokens reduce CER at all four evaluated rates.
  However, the probe is trained on dynamic tokens at those same rates,
  so this is a shared multi-rate representation diagnostic, not a
  symmetric training comparison or a test of unseen-rate generalization.
  Greedy WER remains near one; the frozen external ASR evaluation in
  Appendix~\ref{app:frozen_asr} provides a more interpretable recognition
  measure.

  \subsection{Unseen-Speaker Verification}
  \label{app:speaker_verification}

  The speech representation splits contain disjoint speaker sets:
  251 training speakers and 40 speakers each in development and test.
  A shared probe is trained on dynamic tokens at all four rates using
  mean--standard-deviation pooling, a 256-dimensional normalized
  embedding, and AAM-Softmax supervision on training speakers.
  Checkpoint selection uses mean development EER across dynamic rates.

  Test evaluation uses all unordered pairs of 500 utterances:
  3,440 same-speaker and 121,310 different-speaker trials.
  Scores are cosine similarities. We report EER and normalized
  minDCF with target prior $0.01$ and unit miss/false-alarm costs.

  \begin{table}[t]
  \caption{
  Unseen-speaker verification results, mean $\pm$ standard deviation
  across five seeds. EER is a fraction; minDCF is normalized.
  Bitrates match the corresponding conditions in
  Table~\ref{tab:app_crossrate_ctc_full}.
  }
  \label{tab:app_speaker_full}
  \centering
  \small
  \setlength{\tabcolsep}{7pt}
  \begin{tabular}{@{}clrr@{}}
  \toprule
  \textbf{Depth} & \textbf{Condition}
  & \textbf{EER $\downarrow$}
  & \textbf{minDCF $\downarrow$}\\
  \midrule
  2 & Fixed   & $0.166300\pm0.009085$ & $0.933680\pm0.024685$\\
    & Dynamic & $0.160640\pm0.004255$ & $0.938709\pm0.024173$\\
    & Joint   & $0.160044\pm0.005024$ & $0.941122\pm0.022184$\\
  \midrule
  3 & Fixed   & $0.170977\pm0.004921$ & $0.921212\pm0.027045$\\
    & Dynamic & $0.160026\pm0.005250$ & $0.940062\pm0.020172$\\
    & Joint   & $0.161162\pm0.004663$ & $0.937215\pm0.023602$\\
  \midrule
  4 & Fixed   & $0.178204\pm0.004632$ & $0.932332\pm0.013108$\\
    & Dynamic & $0.161632\pm0.004057$ & $0.949428\pm0.023620$\\
    & Joint   & $0.163372\pm0.004526$ & $0.938332\pm0.022727$\\
  \midrule
  6 & Fixed   & $0.196291\pm0.013163$ & $0.942388\pm0.017924$\\
    & Dynamic & $0.163945\pm0.004219$ & $0.967044\pm0.023801$\\
    & Joint   & $0.169228\pm0.003234$ & $0.959892\pm0.026443$\\
  \bottomrule
  \end{tabular}
  \end{table}

  \begin{table}[t]
  \caption{
  Paired EER differences, mean $\pm$ pointwise 95\% interval half-width.
  Intervals exclude zero at $D=3,4,6$, with 5/5 seed wins for both
  comparisons. At $D=2$, both have 4/5 wins but inconclusive intervals.
  }
  \label{tab:app_speaker_diff}
  \centering
  \small
  \setlength{\tabcolsep}{9pt}
  \begin{tabular}{@{}crr@{}}
  \toprule
  \textbf{Depth}
  & \textbf{Dynamic $-$ Fixed}
  & \textbf{Joint $-$ Fixed}\\
  \midrule
  2 & $-0.005660\pm0.009301$ & $-0.006256\pm0.008753$\\
  3 & $-0.010951\pm0.002664$ & $-0.009815\pm0.001270$\\
  4 & $-0.016573\pm0.004963$ & $-0.014833\pm0.007337$\\
  6 & $-0.032345\pm0.013059$ & $-0.027062\pm0.012910$\\
  \bottomrule
  \end{tabular}
  \end{table}

  EER improves at $D=3,4,6$, but minDCF does not improve uniformly.
  As with CTC, the probe is trained on dynamic conditions, so fixed-token
  evaluation involves a condition shift.

  Joint allocation also exhibits trade-offs relative to acoustic
  allocation: at $D=6$, joint-minus-dynamic EER is
  $+0.005283\pm0.004204$; at $D=4$, joint-minus-dynamic minDCF is
  $-0.011096\pm0.004312$. Both pointwise intervals exclude zero.
  These findings show task- and operating-point-dependent effects,
  rather than uniform improvement from semantic allocation.

  \subsection{FSD50K Environmental-Sound Classification}
  \label{app:fsd50k}

  We use the official FSD50K train/development/test partitions containing
  36,796, 4,170, and 10,231 examples. A shared 200-class multi-label
  probe is trained on fixed and dynamic $D=4$ tokens. Checkpoint
  selection uses mean fixed/dynamic development macro-mAP.

  \begin{table}[t]
  \caption{
  Five-seed FSD50K results. Quality entries are mean $\pm$ standard
  deviation; differences and intervals are paired across seeds.
  }
  \label{tab:app_fsd50k}
  \centering
  \small
  \setlength{\tabcolsep}{4pt}
  \begin{tabular}{@{}lrrrr@{}}
  \toprule
  \textbf{Metric} & \textbf{Fixed} & \textbf{Dynamic}
  & \textbf{Difference} & \textbf{95\% paired CI}\\
  \midrule
  Macro-mAP
  & $0.068051\pm0.000820$
  & $0.069684\pm0.001025$
  & $+0.001633$ & $[+0.001204,+0.002062]$\\
  Micro-mAP
  & $0.201742\pm0.001198$
  & $0.203521\pm0.001463$
  & $+0.001778$ & $[-0.000327,+0.003883]$\\
  Rate (kbps)
  & 6.109820 & 6.025438 & $-0.084382$ & --\\
  \bottomrule
  \end{tabular}
  \end{table}

  Dynamic macro-mAP improves in all five seeds while using
  $1.381\%$ less serialized bitrate. Micro-mAP improves in four
  seeds, but its paired interval includes zero. The supported
  classification gain is therefore specific to macro-mAP.

  \subsection{Artist-Disjoint Music Tagging}
  \label{app:mtg}

  For the MTG-Jamendo downstream probe, we combine the available training
  and validation token sets and repartition them by artist. The new
  train/development/test splits contain 44,518/5,600/5,583 recordings
  from 2,915/309/349 artists, with no artist overlap between these
  probe splits. The vocabulary comprises the 50 most frequent tags
  in the new training split. Each seed trains a shared fixed/dynamic
  multi-label probe.

  This artist-disjoint property applies to downstream probe training.
  The split was constructed after upstream utility training and does
  not establish that every test recording or artist was absent from
  all upstream training data.

  \begin{table}[t]
  \caption{
  Five-seed MTG-Jamendo Top-50 test results at $D=4$.
  Values are means; confidence intervals apply to paired seed
  differences. Dynamic bitrate is $0.2582\%$ lower.
  }
  \label{tab:app_mtg}
  \centering
  \small
  \setlength{\tabcolsep}{6pt}
  \begin{tabular}{@{}lrrrr@{}}
  \toprule
  \textbf{Metric} & \textbf{Fixed} & \textbf{Dynamic}
  & \textbf{Difference} & \textbf{95\% paired CI}\\
  \midrule
  Macro-mAP
  & 0.100553 & 0.100682
  & $+0.000128$ & $[-0.000542,+0.000799]$\\
  Micro-mAP
  & 0.211426 & 0.212286
  & $+0.000859$ & $[-0.000375,+0.002093]$\\
  Rate (kbps)
  & 6.023859 & 6.008305 & $-0.015554$ & --\\
  \bottomrule
  \end{tabular}
  \end{table}

  Macro-mAP improves in four seeds and micro-mAP in all five, but
  both intervals include zero. We report no statistically significant
  difference on these metrics; these comparisons do not establish
  performance equivalence or non-inferiority.

  \subsection{Clotho Audio--Text Retrieval}
  \label{app:clotho}

  Clotho evaluation uses 3,839 training, 1,045 validation, and 1,045
  test audio examples. The test set contains 5,225 captions.
  A shared fixed/dynamic audio--text probe is trained for each seed,
  with selection based on mean validation recall across conditions.

  Mean recall is the unweighted average of audio-to-text and
  text-to-audio R@1, R@5, and R@10. Audio-to-text retrieval counts
  a query as successful when at least one associated caption is
  retrieved within the specified rank.

  \begin{table}[t]
  \caption{
  Clotho test results at $D=4$, averaged across five seeds.
  Intervals are pointwise and unadjusted across retrieval metrics.
  }
  \label{tab:app_clotho}
  \centering
  \small
  \setlength{\tabcolsep}{5pt}
  \begin{tabular}{@{}lrrrr@{}}
  \toprule
  \textbf{Metric} & \textbf{Fixed} & \textbf{Dynamic}
  & \textbf{Difference} & \textbf{95\% paired CI}\\
  \midrule
  Mean recall
  & 0.022960 & 0.023898
  & $+0.000938$ & $[-0.000859,+0.002734]$\\
  A2T R@10
  & 0.042297 & 0.045550
  & $+0.003254$ & $[+0.000043,+0.006464]$\\
  Rate (kbps)
  & 6.025979 & 5.989811 & $-0.036168$ & --\\
  \bottomrule
  \end{tabular}
  \end{table}

  The dynamic mean-recall point estimate is $4.08\%$ higher, with
  three seed wins and $0.6002\%$ lower bitrate, but its interval
  includes zero. A2T R@10 improves in four seeds with one tie and
  is the only individual recall metric whose pointwise interval
  excludes zero. We treat it as a secondary finding, not evidence
  of an overall retrieval improvement.

  \subsection{Summary Across Tasks}
  \label{app:representation_summary}

  The frozen-token results are task dependent. Shared multi-rate probes
  show lower CER and, at $D=3,4,6$, lower speaker EER; a probe trained
  jointly on fixed/dynamic tokens also shows higher FSD50K macro-mAP.
  Speaker minDCF exhibits different trade-offs. The primary MTG and
  Clotho intervals include zero despite positive point estimates and
  lower bitrates. These results support selective benefits of adaptive
  allocation, without establishing universal downstream improvement.

  \section{Subjective Evaluation and External Codec Comparison}
  \label{app:additional_eval}

  The listening study evaluates perceptual changes within the same frozen
  backbone, while the external comparison contextualizes absolute speech
  reconstruction quality. Both use acoustic utterance-level allocation,
  not the causal streaming allocator.

  \subsection{Three-Domain MUSHRA Protocol}
  \label{app:mushra_protocol}

  We conduct a MUSHRA-style study based on the reference-and-anchor
  procedure of ITU-R BS.1534 \citep{itu2015bs1534}.
  The set contains 30 trials: 10 speech, 10 music, and 10 environmental
  examples. Speech excerpts last 5~s; music and environmental excerpts
  last 10~s. All stimuli are stored as 48-kHz mono PCM16.

  Each trial provides a visible reference and four hidden conditions:
  the hidden reference, fixed $D=4$ reconstruction, dynamic $D=4$
  reconstruction, and a low-pass anchor generated with a biquad filter
  at 3.5~kHz. A common peak-scaling factor is applied to all conditions
  within a trial. Hidden-condition labels and trial order are randomized
  when generating the listening set; the same generated set is used
  for all listeners.

  Twenty listeners complete all 30 trials, yielding
  $20\times30\times4=2{,}400$ ratings. All pass the applied
  hidden-reference quality check, so none are excluded.
  For each comparison, ratings are first averaged within each listener
  over the relevant trials, and the resulting listener-level differences
  form the paired statistical observations.

  \subsection{Listening-Set Bitrates}
  \label{app:mushra_rate}

  Table~\ref{tab:app_mushra_rate} reports the container bitrates recorded
  for the source token sequences used to generate listening stimuli.
  The reconstructions are subsequently cropped to the 5- or 10-second
  listening excerpts. These rates therefore describe the source
  representations, not separately encoded containers for the cropped
  playback excerpts.

  \begin{table}[t]
  \caption{
  Mean serialized bitrate of source token sequences used for the
  listening set. Playback excerpts are cropped after reconstruction.
  Percentage changes are computed from the domain mean rates.
  }
  \label{tab:app_mushra_rate}
  \centering
  \small
  \setlength{\tabcolsep}{8pt}
  \begin{tabular}{@{}lrrr@{}}
  \toprule
  \textbf{Domain}
  & \textbf{Fixed (kbps)}
  & \textbf{Dynamic (kbps)}
  & \textbf{Rate change}\\
  \midrule
  Speech      & 6.0567 & 6.0353 & $-0.3528\%$\\
  Music       & 6.0238 & 6.0090 & $-0.2465\%$\\
  Environment & 6.0308 & 5.9900 & $-0.6768\%$\\
  \bottomrule
  \end{tabular}
  \end{table}

  \subsection{MUSHRA Results}
  \label{app:mushra_results}

  The hidden reference and anchor receive overall mean scores of
  $97.05$ and $32.32$, respectively. Fixed and dynamic reconstruction
  receive $68.09$ and $69.03$.

  \begin{table}[t]
  \caption{
  MUSHRA-style results for utterance-level allocation.
  Intervals are paired listener-level 95\% Student-$t$ intervals
  ($df=19$), unadjusted for multiple comparisons.
  }
  \label{tab:app_mushra_full}
  \centering
  \small
  \setlength{\tabcolsep}{7pt}
  \begin{tabular}{@{}lrrrr@{}}
  \toprule
  \textbf{Domain} & \textbf{Fixed} & \textbf{Dynamic}
  & \textbf{Difference} & \textbf{95\% paired CI}\\
  \midrule
  Speech
  & 72.88 & 76.40 & $+3.52$ & $[+2.12,+4.92]$\\
  Music
  & 65.33 & 65.84 & $+0.51$ & $[-0.96,+1.97]$\\
  Environment
  & 66.06 & 64.85 & $-1.22$ & $[-2.71,+0.28]$\\
  Overall
  & 68.09 & 69.03 & $+0.94$ & $[-0.20,+2.07]$\\
  \bottomrule
  \end{tabular}
  \end{table}

  The overall paired difference is $+0.937$ points, with positive
  differences for 15 of 20 listeners. Its 95\% Student-$t$ interval
  is $[-0.195,+2.068]$, the participant-bootstrap interval is
  $[-0.117,+1.950]$, and the paired $t$-test gives $p=0.0994$.

  Speech improves by $3.52$ points with an interval excluding zero.
  Music, environmental audio, and the overall comparison remain
  inconclusive; the environmental point estimate favors fixed coding.
  Thus, the study supports a speech-specific subjective benefit,
  without establishing a general three-domain improvement or
  equivalence in the other domains. Listener-level inference is
  conditional on the selected listening items.

  \subsection{External EnCodec Comparison}
  \label{app:encodec}

  We compare official 24-kHz EnCodec at its nominal 6-kbps,
  eight-codebook operating point with fixed and dynamic $D=4$
  UniAdapt. Evaluation uses all 2,620 LibriSpeech test-clean
  utterances, each truncated to at most 20~s, with the same
  PESQ/STOI evaluation procedure. Audio is converted to mono and
  resampled to each codec's input rate without peak normalization.
  PESQ uses wideband evaluation at 16~kHz.

  UniAdapt uses the three-domain acoustic controller and utterance-level
  allocation. EnCodec rate is measured using its official compression
  interface with \texttt{use\_lm=False}; UniAdapt rate uses its dynamic
  container, including all side information and padding.

  \begin{table}[t]
  \caption{
  Reconstruction on 2,620 test-clean utterances with a common
  20-second duration cap. Values are means over utterances.
  Rates include each system's actual container overhead.
  }
  \label{tab:app_encodec}
  \centering
  \small
  \setlength{\tabcolsep}{8pt}
  \begin{tabular}{@{}lrrr@{}}
  \toprule
  \textbf{System}
  & \textbf{PESQ $\uparrow$}
  & \textbf{STOI $\uparrow$}
  & \textbf{kbps}\\
  \midrule
  Official EnCodec
  & 2.747038 & 0.937710 & 6.112894\\
  UniAdapt fixed $D=4$
  & 1.912894 & 0.870564 & 6.075926\\
  UniAdapt dynamic $D=4$
  & 2.042566 & 0.888684 & 6.043176\\
  \bottomrule
  \end{tabular}
  \end{table}

  Uncertainty is estimated using 10,000 paired utterance-bootstrap
  resamples. Differences in Table~\ref{tab:app_encodec_bootstrap}
  are candidate minus reference; intervals are pointwise.

  \begin{table}[!htbp]
  \caption{
  Paired differences over the same 2,620 utterances.
  Positive PESQ/STOI differences favor the candidate; negative
  rate differences indicate fewer bits per second.
  }
  \label{tab:app_encodec_bootstrap}
  \centering
  \small
  \setlength{\tabcolsep}{6pt}
  \begin{tabular}{@{}lrr@{}}
  \toprule
  \textbf{Comparison / metric}
  & \textbf{Difference}
  & \textbf{95\% bootstrap CI}\\
  \midrule
  Dynamic $-$ Fixed, PESQ
  & $+0.129672$ & $[+0.122993,+0.136169]$\\
  Dynamic $-$ Fixed, STOI
  & $+0.018120$ & $[+0.017752,+0.018490]$\\
  Dynamic $-$ Fixed, kbps
  & $-0.032750$ & $[-0.034445,-0.031083]$\\
  \midrule
  Dynamic $-$ EnCodec, PESQ
  & $-0.704472$ & $[-0.712897,-0.696145]$\\
  Dynamic $-$ EnCodec, STOI
  & $-0.049027$ & $[-0.049624,-0.048433]$\\
  Dynamic $-$ EnCodec, kbps
  & $-0.069718$ & $[-0.072230,-0.067265]$\\
  \bottomrule
  \end{tabular}
  \end{table}

  Dynamic allocation improves PESQ and STOI over the same frozen
  UniAdapt backbone while using $0.5390\%$ less mean serialized
  bitrate; all three paired intervals exclude zero. EnCodec achieves
  higher absolute PESQ and STOI, while UniAdapt dynamic uses
  $1.1405\%$ less mean bitrate than EnCodec.

  The systems differ in architecture, sample rate, training objective,
  quantization, and container format. This comparison therefore
  contextualizes reconstruction quality rather than isolating the
  effect of allocation across architectures.

  \subsection{Overall Evaluation Scope}
  \label{app:evaluation_scope}

  Within the frozen backbone, dynamic allocation improves objective
  speech reconstruction and produces a speech-specific listening
  benefit. Other listening domains and the primary music-tagging
  and retrieval summaries remain statistically inconclusive.
  The external comparison also shows that allocation gains do not
  remove limitations of the underlying codec. The evidence supports
  content-adaptive allocation under explicit serialized budgets,
  without establishing universal downstream gains or reconstruction
  state of the art.

\end{document}